\pdfoutput=1
\documentclass[acmtog,authorversion]{acmart}
\usepackage{booktabs,amsthm} % For formal tables
\usepackage[british]{babel}
\usepackage{subcaption}
\usepackage{graphicx}
\usepackage{enumitem}
\usepackage{interval}

\usepackage[normalem]{ulem}

\newcommand{\y}{\xi}

\newcommand{\q}{\textbf{q}}
\newcommand{\uu}{\textbf{u}}

\usepackage{multirow}
\usepackage{comment}
\usepackage[ruled]{algorithm2e} 

\SetAlFnt{\small}
\SetAlCapFnt{\small}
\SetAlCapNameFnt{\small}
\SetAlCapHSkip{0pt}

\AtBeginDocument{%
  \providecommand\BibTeX{{%
    \normalfont B\kern-0.5em{\scshape i\kern-0.25em b}\kern-0.8em\TeX}}}

\setcopyright{rightsretained}
\acmJournal{TOG}
\acmYear{2022} \acmVolume{1} \acmNumber{1} \acmArticle{1} \acmMonth{1} \acmPrice{1}
\acmDOI{10.1145/3517746}

\begin{document}

\title{OptiTrap: Optimal Trap Trajectories for Acoustic Levitation Displays}

\author{Viktorija Paneva}
\email{vpaneva@acm.org}
\orcid{0000-0002-5152-3077}
\affiliation{%
  \institution{University of Bayreuth}
 % \streetaddress{Universitätsstraße 30}
% \city{Bayreuth}
  \country{Germany}
 % \postcode{95447}
}

\author{Arthur Fleig}
\affiliation{%
  \institution{University of Bayreuth}
%  \streetaddress{Universitätsstraße 30}
%\city{Bayreuth}
  \country{Germany}
 % \postcode{95447}
}
\email{arthur.fleig@uni-bayreuth.de}

\author{Diego Martínez Plasencia}
\affiliation{%
  \institution{University College London}
%  \streetaddress{169 Euston Road}
 %\city{London}
  \country{United Kingdom}
 % \postcode{WC1E 6BT}
 }
\email{d.plasencia@ucl.ac.uk}

\author{Timm Faulwasser}
\affiliation{%
  \institution{TU Dortmund University}
%\city{Dortmund}
  \country{Germany}}
\email{timm.faulwasser@tu-dortmund.de}

\author{J\"{o}rg M\"{u}ller}
\affiliation{%
  \institution{University of Bayreuth}
 % \streetaddress{Universitätsstraße 30}
 %\city{Bayreuth}
  \country{Germany}
 % \postcode{95447}
 }
\email{joerg.mueller@uni-bayreuth.de}

\renewcommand{\shortauthors}{Paneva, Fleig, Martínez, Faulwasser, and M\"{u}ller}

\begin{abstract}
Acoustic levitation has recently demonstrated the ability to create volumetric content by trapping and quickly moving particles along reference paths to reveal shapes in mid-air. 
However, the problem of specifying physically feasible trap trajectories to display desired shapes remains unsolved. 
Even if only the final shape is of interest to the content creator, the trap trajectories need to determine where and when the traps need to be, for the particle to reveal the intended shape.
We propose \emph{OptiTrap}, the first structured numerical approach to compute trap trajectories for acoustic levitation displays. 
Our approach generates trap trajectories that are physically feasible and nearly time-optimal, and reveal generic mid-air shapes, given only a reference path (i.e., a shape with no time information). 
We provide a multi-dimensional model of the acoustic forces around a trap to model the trap-particle system dynamics and compute optimal trap trajectories by formulating and solving a non-linear path following problem. 
We formulate our approach and evaluate it, demonstrating how \emph{OptiTrap} consistently produces feasible and nearly optimal paths, with increases in size, frequency, and accuracy of the shapes rendered, allowing us to demonstrate larger and more complex shapes than ever shown to date. 
\end{abstract}




\maketitle

\section{Introduction}

\begin{figure}[t]
  \includegraphics[width=\columnwidth]{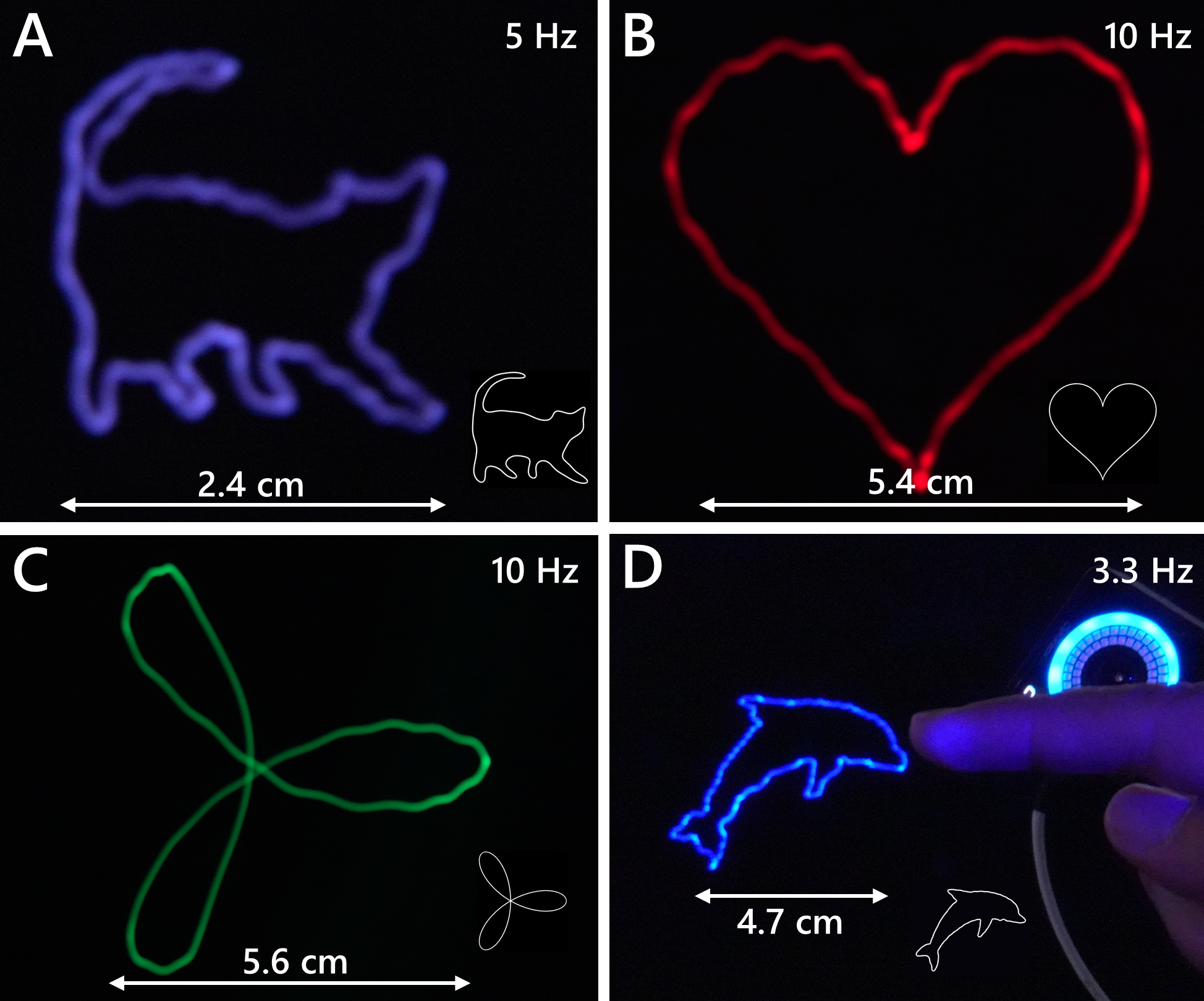}
  \caption{While previously, shapes demonstrated on levitation displays were limited to simple shapes with almost constant curvature \cite{Fushimi19,Hirayama19,Plasencia20}, our approach allows to render generic complex paths. Shapes that have not been demonstrated before include sharp edges as with the heart (B) and dolphin (D), as well as significant changes of the curvature, as with the cat (A) and flower (C) (see Supplementary Video). 
  Photos of the physical particle are made with an exposure time ranging from 0.2 to 1~s, such that each photo shows multiple periods of the orbit.}
  \label{fig:teaser}
  \Description{Four photos labelled A, B, C and D show different graphics generated with the levitation display. Image A shows an outline of a 2.4 centimetres wide cat, rendered at 5 Hertz. Image B shows an outline of a 5.4 centimetres wide heart, rendered at 10 Hertz. Image C shows an outline of a 5.6 centimetres wide three-petal flower, rendered at 10 Hertz. Image D shows an outline of a 4.7 centimetres wide dolphin, rendered at 3.3 Hertz. A human index finger is petting the dolphin. The volumetric dolphin graphics and the index finger are of similar size.}
\end{figure}

\begin{figure*}[!tb]
  \includegraphics[width=2\columnwidth]{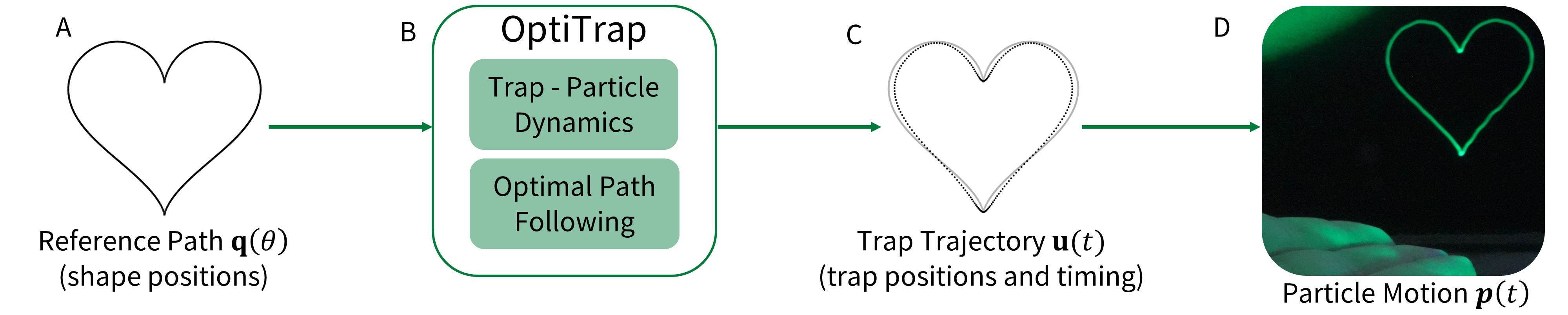}
  \caption{\emph{OptiTrap} is an automated method to compute trap trajectories to reveal generic mid-air shapes on levitation displays. The method accepts a reference path (e.g., the shape of a heart) without any timing information as an input (A). Our approach considers the capabilities of the device and its trap-particle dynamics and combines these with a path following approach (B). The approach produces feasible trap trajectories, describing when and where the traps must be created (C). The resulting particle motion can be presented on the actual device, yielding feasible paths and supporting complex objects with sharp edges and/or significant changes of the curvature (D).}
 \label{fig:teaser_diagram}
 \Description{A flowchart with four elements sequentially connected with flow links. The starting state A is a reference path q of theta, which captures the positions of a reference shape of a heart. A flows to B, the OptiTrap algorithm, composed of the trap-particle dynamics and optimal path following components. The algorithm outputs a trap trajectory u of t, in state C, which contains both the trap positions and timings. An image shows that the trap trajectories do not overlap with the shape positions that were fed to the algorithm in state A. Finally, the ending state D is the particle motion. A photo of the rendered heart shape on the levitation display is shown above a stretched out hand, indicating that the shape is three dimensional and rendered in mid-air.}
\end{figure*}

Acoustic levitation has recently demonstrated the creation of volumetric content in mid-air by exploiting the Persistence of Vision (PoV) effect. 
This is achieved by using acoustic traps to rapidly move single~\cite{Hirayama19} or multiple~\cite{Plasencia20} particles along a periodic reference path, revealing a shape within $0.1s$ (i.e., the integration interval of human eyes~\cite{Bowen74}).

However, the way to define such PoV content remains unsolved. 
That is, there are no automated approaches to compute the \emph{trap trajectories}, i.e., the positioning and timing of the acoustic traps that will allow us to reveal the desired shape. 

Currently, content creators can only rely on trial and error to find physically feasible trap trajectories resembling their intended target shapes, resulting in a time consuming and challenging process. For instance, the creator will need to define the timing of the path (i.e., not only \emph{where} the particle must be, but also \emph{when}). 
Such timing is not trivial and will affect the overall rendering time and the accelerations applied to the particle, which must be within the capabilities of the levitator. 
The way forces distribute around the acoustic trap will also need to be considered to decide where traps must be located.

Considering these challenges is crucial to design feasible trap trajectories, as a single infeasible point along the trajectory typically results in the particle being ejected from the levitator (e.g., approaching a sharp corner too quickly). Thus, while it has been theorised that linear PoV paths of up to 4m and peak speeds of 17m/s are possible~\cite{FushimiLimits}, actual content demonstrated to date has been limited to simple shapes with almost constant curvature along the path and much lower speeds (e.g., 0.72m/s in~\cite{Ochiai14}; 1.2m/s in~\cite{Hirayama19}; 2.25m/s in \cite{Plasencia20}, combining six particles). 

In this paper, we present \emph{OptiTrap}, the first structured numerical approach to compute trap trajectories for acoustic levitation displays.
\emph{OptiTrap} automates the definition of levitated PoV content, computing physically feasible and nearly time-optimal trap trajectories given only a reference path, i.e., the desired shape. 
As shown in Figure~\ref{fig:teaser}, this allows for larger and more complex shapes than previously demonstrated, as well as shapes featuring significant changes in curvature and/or sharp corners.

Our approach is summarised in Figure~\ref{fig:teaser_diagram}. 
\emph{OptiTrap} assumes only a generic reference path $\q(\theta)$ as an input, with no temporal information (see~Figure~\ref{fig:teaser_diagram}(A)). \emph{OptiTrap} formulates this as a path following problem (see~Figure~\ref{fig:teaser_diagram}(B)), computing the optimum timing in which a particle can traverse such path. Our formulation considers the \emph{Trap-Particle Dynamics} of the system, using a 3D model of acoustic forces around a trap. This results in a non-invertible model, which cannot exploit differential flatness~\cite{Fliess95a}, on which most path following approaches rely. We instead provide a coupling stage for the dynamics of our system and numerically invert the system. This approach produces a trap trajectory $\uu(t)$ (Figure~\ref{fig:teaser_diagram}(C)) that results in the intended and nearly time-optimal particle motion. That is, $\uu(t)$ defines the positions and timing of the traps that cause the particle to reveal the target shape $\q(\theta)$, according to the capabilities of the actual device, cf.~Figure~\ref{fig:teaser_diagram}(D).  

In summary, we contribute the first structured numerical approach to compute physically feasible and nearly time-optimal trap trajectories for levitation displays, given only a reference path that the particle should follow. As a core contribution, we provide a theoretical formulation of the problem in terms of path following approaches, allowing optimum timing and device properties (e.g., trap-particle dynamics) to be jointly considered. All the details of the approach we propose are provided in Section~\ref{sec:opt}.

We illustrate the potential of our approach by rendering shapes featuring straight lines, sharp corners, and complex shapes, such as those in Figure~\ref{fig:teaser}. We then provide an experimental validation of our approach, demonstrating increases of up to 563\% in the size of rendered objects, up to 150\% in the rendering frequency, and improvements in accuracy (e.g., shape revealed more accurately). While the baseline shapes we compare against require trial and error to determine optimal sizes or frequencies, our approach always yields feasible paths (i.e., working in at least 9 out of 10 attempts) and makes consistent use of accelerations very close to (but not exceeding) the maximum achievable by the device, independently of the target shape.

These features allow \emph{OptiTrap} to render complex objects involving sharp edges and significant changes in curvature that have never been demonstrated before. Even more importantly, it provides a tool to systematically explore the range of contents that levitation displays can create, as a key step to exploit their potential.

%----------------------------------------------------------------------------------------------
\section{Background and Challenges}
\label{sec:prob_stat}

\begin{figure*}[!tb]
  \includegraphics[width=2\columnwidth]{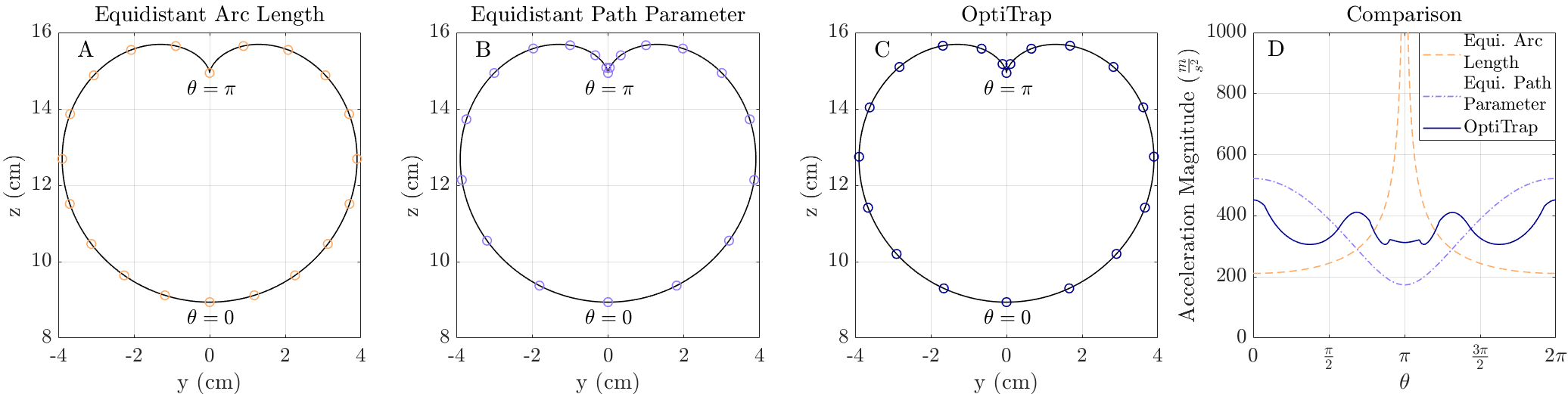}
  \caption{	  Three different path timings for the cardioid shape with fixed traversal time: (A) equidistant sampling of the arc length, (B) equidistant sampling of the path parameter~$\theta$, and (C) \emph{OptiTrap}. 
	  The corresponding acceleration magnitudes are depicted in (D). 
	  Strategy (A) is physically infeasible due to infinite required acceleration at the corner ($\theta=\pi$). 
	  Strategy (B) does not share this problem, but requires careful parametrization of the path, while \emph{OptiTrap} (C) yields the timing automatically.}
  \label{fig:timings}
\Description{The figure consists of four subplots. The first three plots show a cardioid shape in two dimentional space. The x-axis denotes the horizontal spatial component from -4 to 4 centimetres and the y-axis the vertical from 8 to 16 centimetres, both at increments of two centimetres. Subplot D shows the acceleration magnitude resulting from the timing strategies from A to C, on a scale from 0 to 1000 metres per seconds squared as a function of the path parameter theta, going from 0 to two pi. The most notable difference in the timing strategies occurs at the sharp corner of the cardioid shape, where the equidistant arc length strategy results in infinite acceleration. }
\end{figure*}

Our goal is is to minimise complexity for content creators. Thus, given a reference path (i.e., a shape defined by the content creator), our approach must produce physically feasible trap trajectories, which accurately reveal the desired path, while making optimal use of the capabilities of the device. 

We formalise our content as a reference path $Q$, provided by the content creator without any preassigned time information.
It is an explicitly parametrized curve
\begin{equation} \label{eq:path}
    Q:=\{\y \in \mathbb{R}^3 \mid \theta \in [\theta_{0},\theta_{f}] \mapsto \q(\theta) \},
\end{equation}
where $\q\colon\mathbb{R} \rightarrow \mathbb{R}^3$, as later justified in Section~\ref{ssec:OPT}, must be a twice continuously differentiable function with respect to~$\theta$, which is called the path parameter.
Increasing values of~$\theta$ denote forward movement along the path~$Q$.
The starting point on the path is $\q(\theta_0)$, while $\q(\theta_f)$ marks the end of the path. 
For periodic paths, such as a particle cyclically revealing a PoV shape, $\q(\theta_0) = \q(\theta_f)$ and $\dot{\q}(\theta_0) = \dot{\q}(\theta_f)$ hold. 
Please note that while splines would provide a generic solution to this parametrization (i.e., twice continuously differentiable fitting the points in $Q$), any other parametric functions can be used.
For example, taking $\theta\in[0, 2\pi]$ and $r>0$, a cardioid as in Figure~\ref{fig:timings}, can be described by: 
\begin{equation}\label{eq:cardioid}
	\q(\theta)=(0, r\sin(\theta)(1+\cos(\theta)), -r\cos(\theta)(1+\cos(\theta))+r)^\top.
\end{equation}

Whatever the approach used, revealing such reference paths typically involves rapid particle movements, as PoV content must be revealed within about $0.1s$~\cite{Bowen74}) and must consider the feasibility of the path (i.e., the trap-particle dynamics, the topology of the trap, and the capabilities of the levitator). This results in the following two main challenges for \emph{OptiTrap}: 1) determining the optimal timing for the particle revealing the path; and 2) computing the trap positions generating the required forces on the particle. 

\subsection{Challenge 1: Determining an Optimal Path Timing} 
The reference path $Q$ defines the geometry but not the timing (i.e., it describes \emph{where} the particle needs to be, but not \emph{when} it needs to be there). Our approach must compute such a timing, and the strategy followed will have important implications on the velocity and accelerations applied to the particle and, hence, on the physical feasibility of the content. 

Different timing strategies and their effects on the feasibility can be illustrated using the cardioid example from~\eqref{eq:cardioid}. Figure~\ref{fig:timings} depicts various timing strategies applied to this shape, all of them traversing the same path in the same overall time. 

A straight-forward approach would be to sample equidistantly along the path, as shown in Figure~\ref{fig:timings}(A). This works well for lines and circles, where a constant speed can be maintained, but fails at sudden changes in curvature due to uncontrolled accelerations (see the acceleration at the corner of cardioid in Figure~\ref{fig:timings}(D), at $\theta=\pi$).

The second example in Figure~\ref{fig:timings}(B) shows a simple equidistant sampling on the path parameter $\theta$. As shown in Figure~\ref{fig:timings}(D), this results in areas of strong curvature naturally involving low accelerations, and can work particularly well for low-frequency path parametrizations in terms of sinusoidals (e.g., \cite{FushimiLimits} showed such sinusoidal timing along straight paths, to theorise optimum/optimistic content sizes). In any case, the quality of the result obtained by this approach will vary depending on the specific shape and parametrization used.

As an alternative, the third example shows the timing produced by our approach, based on the shape and capabilities of the device (i.e., forces it can produce in each direction). Please note how this timing reduces the maximum accelerations required (i.e., retains them below the limits of the device), by allowing the particle to travel faster in parts that were unnecessarily slow (e.g., Figure~\ref{fig:timings}(D), at $\theta=\pi$). For the same maximum accelerations (i.e., the same device), this timing strategy could produce larger shapes or render them in shorter times, for better refresh rates. This illustrates the impact of a careful timing strategy when revealing any given reference path.  
We address this challenge in Section~\ref{ssec:OPT}.

\subsection{Challenge 2: Computation of Trap Positions}
In addition to the timing, the approach must also compute the location of the traps. Previous approaches \cite{Hirayama19, Plasencia20, FushimiLimits} placed the traps along the shape to be presented, under the implied assumption that the particle would remain in the centre of such trap. 

However, the location where the traps must be created almost never matches the location of the particle. Acoustic traps feature (almost) null forces at the centre of the trap and high restorative forces around them \cite{Marzo15}. As such, the only way a trap can accelerate a particle is by having it placed at a distance from the centre of the trap. Such trap-particle distances were measured by \cite{Hirayama19} to assess performance achieved during their speed tests. Distortions related to rendering fast moving PoV shapes were also shown in \cite{Fushimi19}. However, none of them considered or corrected for such displacements. 

Please note the specific displacement between the trap and the particle will depend on several factors, such as the particle acceleration required at each point along the shape, as well as the trap topology (i.e., how forces distribute around it). No assumptions can be made that the traps will remain constrained to positions along or tangential to the target shape.
We describe how our approach solves this challenge in Section~\ref{ssec:extract-trap-trajectory}.

%-----------------------------------------------------------------------------

\section{Related Work}
\emph{OptiTrap} addresses the challenges above by drawing from advances in the fields of acoustic levitation and control theory, which we review in this section.

\subsection{Acoustic Levitation}

Single frequency sound-waves were first observed to trap dust particles in the lobes of a standing wave more than 150 years ago~\cite{Stevens1899}. 
This has been used to create mid-air displays with particles acting as 3D voxels~\cite{Ochiai14,Omirou15,Omirou16, Sahoo16}, but such standing waves do not allow control of individual particles. 
Other approaches have included Bessel beams~\cite{Norasikin19}, self-bending beams~\cite{Norasikin18}, boundary holograms~\cite{Inoue19} or near-field levitation~\cite{NearFieldLevitation}.  

However, most display approaches have relied on the generic levitation framework proposed by~\cite{Marzo15}, combining a focus pattern and a levitation signature. 
Although several trap topologies (i.e., twin traps, vortex traps or bottle beams) and layouts (i.e., one-sided, two-sided, v-shape) are possible, most displays proposed have adopted a top-bottom levitation setup and twin traps, as these result in highest vertical trapping forces. 
This has allowed individually controllable particles/voxels~\cite{Marzo19} or even particles attached to other props and projection surfaces for richer types of content~\cite{Morales19, FenderArticulev}. 
The use of single~\cite{Fushimi19,Hirayama19} or multiple~\cite{Plasencia20} fast moving particles have allowed for dynamic and free-form volumetric content, but is still limited to small sizes and simple vector graphics~\cite{FushimiLimits}. 

Several practical aspects have been explored around such displays, such as selection~\cite{Freeman18} and manipulation techniques~\cite{Bachynskyi18}, content detection and initialisation~\cite{FenderArticulev} or collision avoidance~\cite{Reynal20}. 

However, no efforts have been made towards optimising content considering the capabilities (i.e., the dynamics) of such displays, particularly for challenging content such as the one created by PoV high-speed particles. 
\cite{Hirayama19} showed sound-fields must be updated at very high rates for the trap-particle system to engage in high accelerations, identifying optimum control for rates above 10kHz. 
However, they only provided a few guidelines (e.g., maximum speed in corners, maximum horizontal/vertical accelerations) to guide the definition of the PoV content. 
\cite{FushimiLimits} provided a theoretical exploration of this topic, looking at maximum achievable speeds and content sizes, according to the particle sizes and sound frequency used. 
While the dynamics of the system were considered, these were extremely simplified, using a model of acoustic trap forces and system dynamics that are only applicable for oscillating recti-linear trajectories along the vertical axis of the levitator.
\cite{Paneva20} proposed a generic system simulating the dynamics of a top-bottom levitation system. 
This operates as a forward model simulating the behaviour of the particle given a specific path for the traps, but unlike our approach it does not address the inverse problem. 
Thus, \emph{OptiTrap} is the first algorithm allowing the definition of PoV content of generic shapes, starting only from a geometric definition (i.e., shape to present, no timing information) and  optimising it according to the capabilities of the device and the dynamics of the trap-particle system.   

\subsection{Path Following and Optimal Control} \label{sec:PFopt}
The particle in the trap constitutes a dynamical system, which can be controlled by setting the location of the trap. 
As such, making the particle traverse the reference path can be seen as an optimal control problem (OCP).

In the context of computer graphics, optimal control approaches have been studied particularly in the areas of physics-based character animation~\cite{geijtenbeek12animation} and aerial videography~\cite{nageli2017real}. 
Such OCPs can be considered as function-space variants of nonlinear programs, whereby nonlinear dynamics are considered as equality constraints. In engineering applications, OCPs are frequently solved via direct discretization~\cite{Stryk93,Bock84}, which leads to finite-dimensional nonlinear programs. 
In physics-based character animation, such direct solution methods
are known as spacetime constraints~\cite{witkin88spacetime, rose96spacetime}. 

If framed as \emph{a levitated particle along a given geometric reference path (i.e., the PoV shape)}, the problem can be seen as an instance of a path following problem~\cite{Faulwasser12}.
Such problems also occur in aerial videography, when a drone is to fly along a reference path~\cite{nageli2017real, roberts16generating}. 
In order to yield a physically feasible trajectory, it is necessary to adjust the timing along that trajectory. This can be done by computing feed-forward input signals~\cite{Faulwasser12, roberts16generating} or, if the system dynamics and computational resources allow, using closed-loop solutions, such as Model Predictive Control in~\cite{nageli2017real}. 

However, optimal control of acoustically levitated particles cannot be approached using such techniques. All of the above approaches require {\em differential flatness}~\cite{Fliess95a}.
That is, the underlying system dynamics must be invertible, allowing for the problem to be solved by projecting the dynamics of the moving object onto the path manifold~\cite{Nielsen08a}.
This inversion approach also enables formulating  general path following problems in the language of optimal control, encoding desired objectives (e.g., minimum time, see~\cite{Shin85a,  Verscheure09b, ifat:faulwasser14b}).

As discussed later in Section~\ref{ssec:model-acoustic-forces}, the distribution of forces around the acoustic trap (i.e., our system dynamics) are not invertible, requiring approaches that have not been widely developed in the literature. We do this by coupling the non-invertible particle dynamics into a virtual system, solvable with a conventional path following approach. We then use these coupling parameters and our model of particle dynamics to solve for the location of the traps. 

%-----------------------------------------------------------------------------
\section{Optimal Control for Levitation Displays}\label{sec:opt}

This section provides a description of our \emph{OptiTrap} approach, which automates the definition of levitated content, computing physically feasible and nearly time-optimal trap trajectories using only a reference path as an input. 
In Section~\ref{ssec:model-acoustic-forces}, we first describe our specific hardware setup and the general mathematical framework, then we consider existing models of the trap-particle dynamics (Subsection~\ref{sssec:exsting-models}), before introducing our proposed model (Subsection~\ref{sssec:our-model}).
Next, we describe the two stages in our algorithm, which match the challenges identified in Section~\ref{sec:prob_stat}. 
That is, Section~\ref{ssec:OPT} describes the computation of optimum timing, approached as an optimal path following problem, while Section~\ref{ssec:extract-trap-trajectory} explains how to compute the trap locations. 
Please note that Section~\ref{sec:opt} focuses on the general case of presenting levitated content, that is, particles cyclically traversing a reference path (shape) as to reveal it. 
Other cases, such as a particle accelerating from rest as to reach the initial state (i.e., initial position and speed) required to render the content can be easily derived from the general case presented here, and are detailed in the Supplementary Material S1. 

\begin{figure}[b]
  \includegraphics[width=\columnwidth]{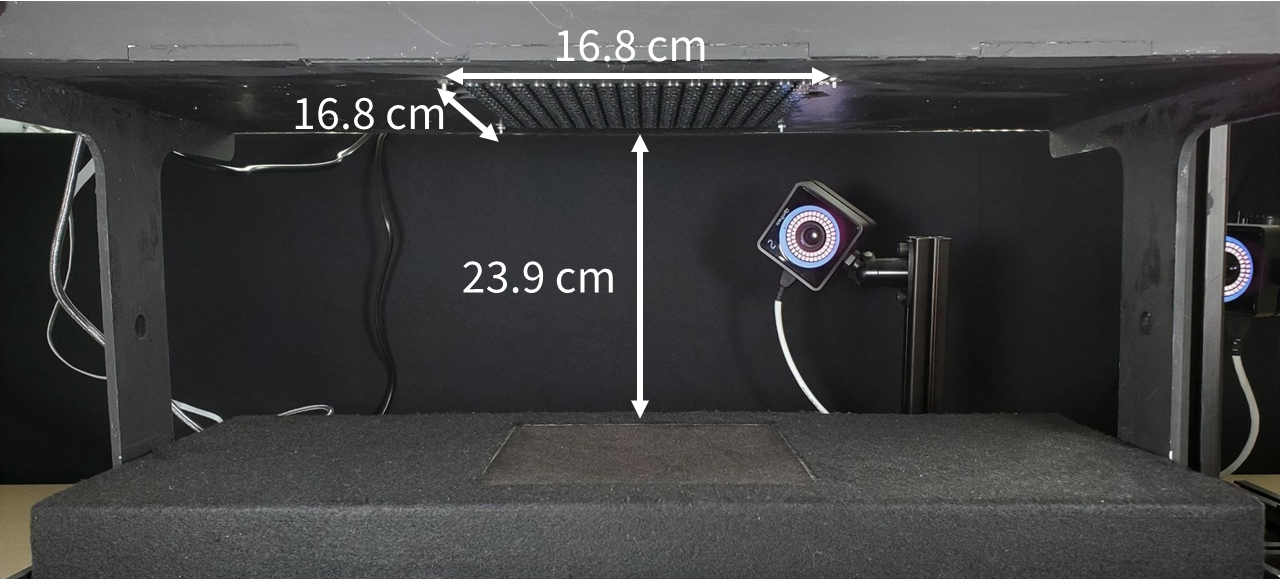}
  \caption{Overview of the components in our setup. We used two opposed arrays of transducers at a distance of 23.9 cm and an OptiTrack system to track the position of levitated particles in real time.  }
  \label{fig:setup}
\Description{A photo of equipment. Two square array boards of transducers with dimension 16.8 centimetres are aligned to exactly face each other at a vertical distance of 23.9 centimetres. Two optical motion cameras mounted on the side, are faced towards the volume between the boards.}
\end{figure}

\subsection{Modelling the Trap-Particle Dynamics}\label{ssec:model-acoustic-forces}

We start by describing the model of the trap-particle dynamics used for our specific setup, shown in Figure \ref{fig:setup}. This setup uses two opposed arrays of 16×16 transducers controlled by an FPGA and an OptiTrack tracking system (Prime 13 motion capture system at a frequency of 240Hz). The design of the arrays is a reproduction of the setup in~\cite{Morales21}, modified to operate at 20Vpp and higher update rates of 10kHz. The device generates a single twin-trap using the method described in \cite{Hirayama19}, allowing for vertical and horizontal forces of $4.2\cdot10^{-5}$N and $2.1\cdot10^{-5}$N, respectively, experimentally computed using the linear speed tests in \cite{Hirayama19} (i.e., 10cm paths, binary search with 9 out 10 success ratios, with particle mass $m\approx 0.7\cdot10^{-7}$ kg). 
Note the substantial difference between the maximum forces in the vertical and horizontal directions. Our approach will need to remain aware of direction as this determines maximum accelerations.

In general, the trap-particle dynamics of such system can be described by simple Newtonian mechanics, i.e.,
\begin{equation}\label{eq:system}
	m\ddot{\textbf{p}}(t) = F(\textbf{p}(t),\dot{\textbf{p}}(t),\textbf{u}(t)).
\end{equation}
Here, $\textbf{p}(t)=(p_x, p_y, p_z)^\top \in \mathbb{R}^3$ represents the particle position in Cartesian $(x,y,z)$ coordinates at time $t\in\mathbb{R}_0^+$, and $\dot{\textbf{p}}(t)$ and $\ddot{\textbf{p}}(t)$ are the velocity and acceleration of the particle, respectively. The force acting on the particle is mostly driven by the acoustic radiation forces and, as such, drag and gravitational forces can be neglected \cite{Hirayama19}. Therefore, the net force acting on the particle will depend only on $\textbf{p}(t)$ and on the position of the acoustic trap at time $t$ denoted by $\uu(t)=(u_x(t),u_y(t),u_z(t))^\top\in\mathbb{R}^3$. 

As a result, our approach requires an accurate and ideally invertible model of the acoustic forces delivered by an acoustic trap. That is, we seek an accurate model predicting forces at any point around a trap only in terms of $\textbf{u}(t)$ and $\textbf{p}(t)$. 
Moreover, an invertible model would allow us to analytically determine where to place the trap as to produce a specific force on the particle given the particle location.

For spherical particles considerably smaller than the acoustic wavelength and operating in the far-field regime, such as those used by our device, the acoustic forces exerted can be modelled by the gradient of the Gor'kov potential~\cite{Bruus2012}. This is a generic model, suitable to model acoustic forces resulting from any combination of transducer locations and transducer activations, but it also depends on all these parameters, making it inadequate for our approach. 

Our case, using a top-bottom setup and vertical twin traps is much more specific. This allows for simplified analytical models with forces depending only on the relative position of the trap and the particle, which we compare to forces as predicted from the Gor'kov potential in Figure~\ref{fig:modelForces}. 

\subsubsection{Existing Models of the Trap-Particle Dynamics}\label{sssec:exsting-models}

\begin{figure}[!tb]
  \includegraphics[width=\columnwidth]{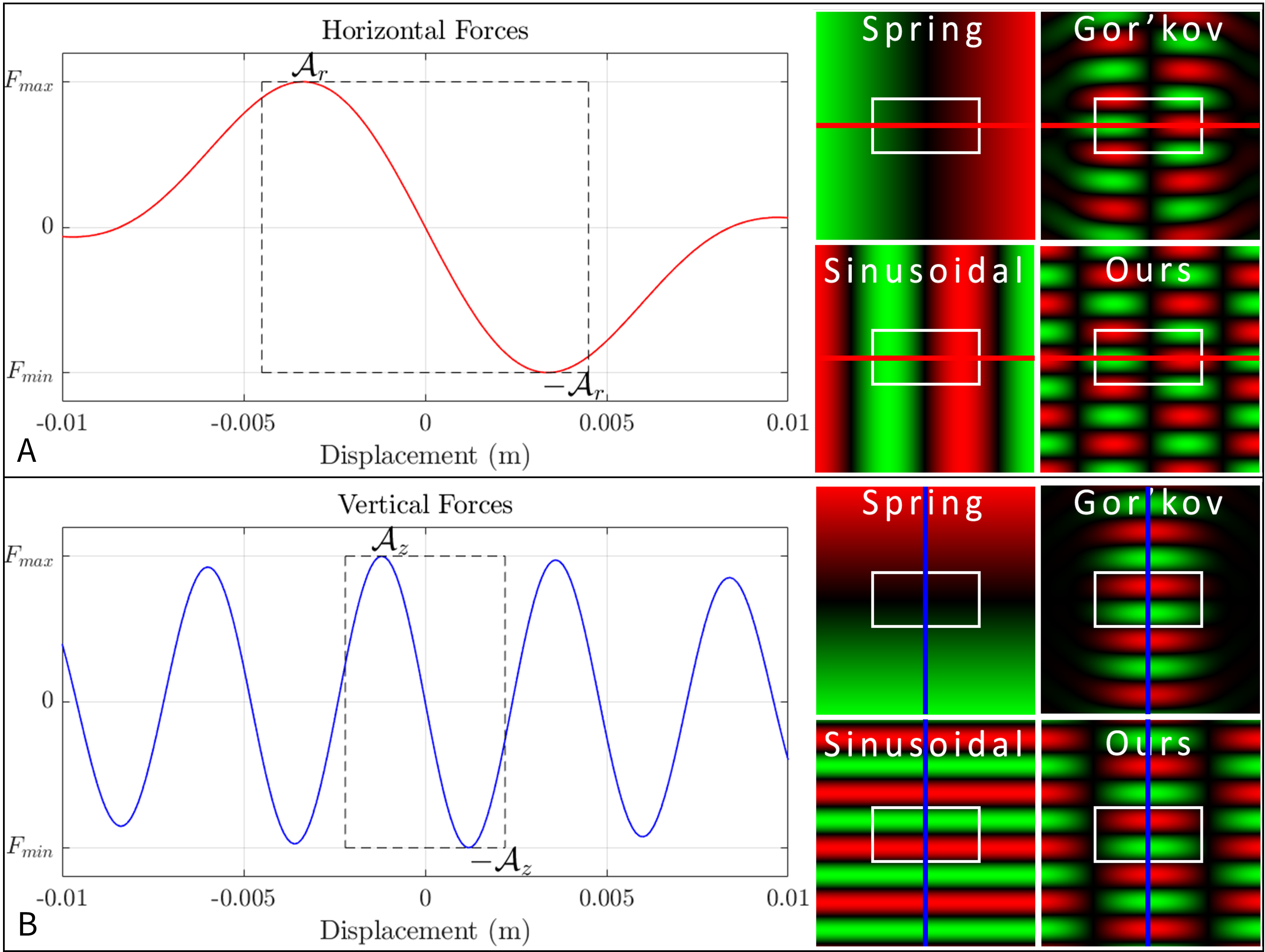}
  \caption{Analytical models of horizontal (A) and vertical (B) acoustic forces around a trap. The plots on the left show forces along the main axes X and Z, and analytical models provide a good fit. The plots on the right show horizontal (A) and vertical (B) forces across 2D slices through the trap centre along the X and Z axes. Spring and Sinusoidal models fail to predict forces outside the main axes, while our proposed model provides accurate reconstruction within the region of interest (highlight).}
  \label{fig:modelForces}
\Description{Two subplots A and B showing the acoustic force on the y-axis going from F maximum to F minimum, as a function of the displacement between the acoustic trap and the particle, ranging from -0.01 to 0.01 metres. }
\end{figure}

Simple \emph{spring} models have been used extensively \cite{Paneva20, Fushimi18Nonlin, FushimiLimits}, modelling trapping forces according to a \emph{stiffness} parameter $\mathcal{K}_i$, that is, with forces being proportional to the distance of the particle to the centre of the trap: 
\begin{equation*}
	F_i(\textbf{p},\textbf{u}) := \mathcal{K}_i \cdot |u_i-p_i|, \quad i \in \{x,y,z\}.
\end{equation*}
Such models are usually refined by providing specific stiffness values for each dimension, but they are only suitable for particles remaining in close proximity to the centre of the trap, i.e., the linear region near the centre of the trap. As a result, \emph{spring} models are only accurate for systems moving particles slowly, requiring low acceleration and forces (so that particles remain within the linear region).

\emph{Sinusoidal} models have been proposed as an alternative~\cite{Fushimi18Nonlin,FushimiLimits}, providing accurate fitting from the centre of the trap to the peaks of the force distribution in Figure~\ref{fig:modelForces}, according to the peak trapping force $\mathcal{A}_i$ and characteristic frequency $\mathcal{V}_i$: 
\begin{equation*}
	F_i(\textbf{p},\textbf{u}) := \mathcal{A}_i \cdot \sin(\mathcal{V}_i \cdot (u_i-p_i)), \quad i \in \{x,y,z\},
\end{equation*}

However, both of these models (i.e., \emph{spring} and \emph{sinusoidal}) are only suitable for particles placed along one of the main axes of the acoustic trap, not for particles arbitrarily placed at any point around it.
This is illustrated on the right of Figure~\ref{fig:modelForces}, which provides an overview of how forces distribute on a horizontal and vertical 2D plane around a trap, according to each model (i.e., Gor'kov, \emph{spring}, \emph{sinusoidal}, and \emph{Ours}, detailed in the next subsection). 

Please note how the three previous models show good matching along the horizontal and vertical axes (represented as blue and red lines). 
However, the \emph{spring} and \emph{sinusoidal} models are of one-dimensional nature (e.g., force $F_x$ only depends on X distance $(u_x-p_x)$) and become inaccurate at points deviating from the main axes. 

\subsubsection{Our Model} \label{sssec:our-model}

The complexity of the force distribution modelled by the Gor'kov potential increases as the distance to the centre of the trap increases. 
However, we only need to derive a model allowing us to predict where to place the trap to produce a specific force on the particle. 
This allows us to limit our considerations to the region corresponding to the peaks designated by $\mathcal{A}_r$ and $\mathcal{A}_z$ in Figure~\ref{fig:modelForces}. 

For points within this region, forces around twin traps distribute in a mostly \emph{axis-symmetric} fashion, which can be approximated as:  
\begin{subequations}\label{eq:FrFz}
	\begin{align}
		F_r(\textbf{p},\textbf{u}) :=\ & \mathcal{A}_r \cdot \cos\left(\mathcal{V}_z \cdot {(u_z-p_z)}\right) \cdot \\
		 &\sin\left(\mathcal{V}_{xr} \cdot {\sqrt{(u_x-p_x)^2+(u_y-p_y)^2}}\right), \notag \\
		F_z(\textbf{p},\textbf{u}) :=\  & \mathcal{A}_z \cdot \sin\left(\mathcal{V}_z \cdot (u_z-p_z)\right) \cdot \\ 
		&  \cos\left(\mathcal{V}_{zr} \cdot \sqrt{(u_x-p_x)^2+(u_y-p_y)^2}\right), \notag 
	\end{align}
\end{subequations}
where $\mathcal{A}_r, \mathcal{A}_z$ represent peak trapping forces along the radial and vertical directions of the trap, respectively. 
$\mathcal{V}_z$, $\mathcal{V}_{xr}$, $\mathcal{V}_{zr}$ represent characteristic frequencies of the sinusoidals describing how the forces evolve around the trap. 
The resulting forces can be converted into acoustic forces in 3D space from these cylindrical coordinates, with azimuth $\phi = \arctan ((u_y-p_y)/(u_x-p_x))$: 
\begin{equation} \label{eq:Facou}
	F(\textbf{p},\textbf{u}) = \begin{pmatrix}
		F_x(\textbf{p},\textbf{u}) \\
		F_y(\textbf{p},\textbf{u}) \\
		F_z(\textbf{p},\textbf{u})
	\end{pmatrix} := \begin{pmatrix}
		F_r(\textbf{p},\textbf{u}) \cos\phi \\
		F_r(\textbf{p},\textbf{u}) \sin\phi \\
		F_z(\textbf{p},\textbf{u})
	\end{pmatrix}.
\end{equation}

We validated our model by comparing its accuracy against the forces predicted by the gradient of the Gor'kov potential. More specifically, we simulated 729 single traps homogeneously distributed across the working volume of our levitator (i.e., 8x8x8cm, in line with \cite{Hirayama19, Fushimi19}), testing 400 points around each trap for a total of 400x729 force estimations. 

\begin{table}[t]
\caption{Fit parameters for the Spring, Sinusoidal and Axis-symmetric models and the respective average relative errors when compared to Gor'kov.}
\centering
\begin{tabular}{ |c|c|c|c|c|c|c|} 
  \hline 
  Model & \multicolumn{2}{|c|}{Spring}& \multicolumn{2}{|c|}{Sinusoidal}& \multicolumn{2}{|c|}{Axis-symmetric}
\\
  \hline 
  \multirow{6}{*}{Fit}&$\mathcal{K}_x$&-0.0071&$\mathcal{A}_x$ &0.00009 &$\mathcal{A}_r$ & 0.0004636\\
 &$\mathcal{K}_y$ & -0.0071&$\mathcal{A}_y$ &0.00009 &$\mathcal{A}_z$ & 0.0002758\\
  &$\mathcal{K}_z$&-0.94&$\mathcal{A}_z$ &-0.0019 &$\mathcal{V}_z$ & 1307.83\\
 & & &$\mathcal{V}_x$ &-68.92 &$\mathcal{V}_{xr}$ & -476.49\\
 &&&$\mathcal{V}_y$ &-68.92  &$\mathcal{V}_{zr}$ & 287.87\\
 & & &$\mathcal{V}_z$ &1307.83 &&\\
  \hline 
Error & \multicolumn{2}{|c|}{63.3\%}& \multicolumn{2}{|c|}{35.9\%}& \multicolumn{2}{|c|}{4.3\%}
\\
  \hline 
\end{tabular}
\label{table:fit_parameters}
\end{table}

Table~\ref{table:fit_parameters} summarises the error distribution achieved by these three models when compared to Gor'kov, showing an average relative error as low as 4\% for our model and much poorer fitting for the other two models. Full details and raw data used in this validation can be found in the Supplementary Material S2. 

As a final summary, this results in a model for our trap-particle dynamics that only depends on $\textbf{p}$ and $\uu$ and which fits the definition of a second-order ordinary differential equation:
\begin{equation}\label{eq:system_acoustic}
	m\ddot{\textbf{p}}(t) = F(\textbf{p}(t),\textbf{u}(t)).
\end{equation}
While the resulting model is accurate (4\% relative error), we note that it is not invertible, which will complicate the formulation of our solution approach. For instance, for a particle at $\textbf{p}=(0,0,0)$, force $F(\textbf{p},\textbf{u}) = (0,0, \mathcal{A}_z \cdot \cos\left(\mathcal{V}_{zr} \cdot x \right))$ can be obtained with either $\textbf{u}= (x, 0, \pi/(2\mathcal{V}_z))$ or $\textbf{u}= (-x, 0, \pi/(2\mathcal{V}_z))$.

\subsection{Open-Loop Optimal Path Following} \label{ssec:OPT}
This section computes the timing for the particle, so that it moves along the given reference path $Q$ from~\eqref{eq:path} in minimum time and according to the dynamics of the system. We approach this as an open-loop (or feed-forward) path following problem, as typically done in robotics~\cite{Faulwasser12}. That is, we design an optimal control problem that computes the timing $t \mapsto\theta(t)$ as to keep the levitated particle on the prescribed path, to traverse the path in optimum (minimum) time while considering the trap-particle dynamics (i.e., minimum \emph{feasible} time). 

Our non-invertible model of particle dynamics calls for a more complex treatment of the problem, which we split in two parts. 
In the first part we derive a virtual system for the timing law, which we describe as a system of first-order differential equations, solvable with traditional methods.
In the second part we couple the timing law with 
our non-invertible dynamics, introducing auxiliary variables that will later enable pseudo-inversion (i.e., to compute trap placement for a given force) as described in Section~\ref{ssec:extract-trap-trajectory}. 

\subsubsection{Error Dynamics and the Timing Law.}\label{sssec:OPT_timing}

The requirement that the particle follows the path $Q$ exactly and for all times means that the deviation from the path equals zero for all $t\in\mathbb{R}^+_0$, i.e., 
\[
\textbf{e}(t) := \textbf{p}(t) -\textbf{q}(\theta(t)) \equiv 0.
\]
If the path deviation $\textbf{e}(t)$ is $0$ during the whole interval $[t_0, t_1]$, this implies that the time derivatives of $\textbf{e}(t)$ also have to vanish on $(t_0, t_1)$.
Considering this (i.e., $\dot e(t)\equiv 0$ and $\ddot e(t)\equiv 0$) results in:
\begin{subequations}\label{eq:dpath_para}
\begin{align}
    \textbf{p}(t) &=\textbf{q}(\theta(t)), \label{eq:dpath_dpara0} \\
    \dot{\textbf{p}}(t)&=\dot{\textbf{q}}(\theta(t))=\frac{\partial \textbf{q}}{\partial \theta}\dot{\theta}(t),\\
    \ddot{\textbf{p}}(t)&= \ddot{\textbf{q}}(\theta(t))=\frac{\partial^2 \textbf{q}}{\partial \theta^2}\dot{\theta}(t)^2+\frac{\partial \textbf{q}}{\partial \theta}\ddot{\theta}(t). \label{eq:dpath_dpara_2}
\end{align}
\end{subequations}

Thus, for particles on the path~$Q$, system dynamics of the form~\eqref{eq:system_acoustic}, and provided we are able to express $\textbf{u}$ as a function of $\textbf{p}$ and $\ddot{\textbf{p}}$ (i.e., the system inversion described in Section~\ref{sec:PFopt}), the
position, the velocity, and the acceleration of the particle can be expressed via $\theta, \dot\theta, \ddot\theta$, respectively. 
For details, a formal derivation, and tutorial introductions we refer to~\cite{Faulwasser12,epfl:faulwasser15c}.

Observe that in~\eqref{eq:dpath_para} the partial derivatives $\frac{\partial^2 \textbf{q}}{\partial \theta^2}$ and $\frac{\partial \textbf{q}}{\partial \theta}$, as well as $\dot\theta$ and~$\ddot\theta$ appear. 
This leads to two additional observations: 
First, the parametrisation $\q$ from~\eqref{eq:path} should be at least twice continuously differentiable with respect to $\theta$, so that the partial derivatives are well-defined.
This justifies our constraint in Section~\ref{sec:prob_stat}. 
Please note this does not rule out corners in the path~$Q$, as shown by the parametrisation~\eqref{eq:cardioid} of the cardioid.
Second, the time evolution of $\theta$ should be continuously differentiable, as otherwise large jumps can occur in the acceleration. 

To avoid said jumps, we
generate the timing $t\mapsto \theta(t)$ via the double integrator
\begin{equation}\label{eq:timingLaw}
	\ddot\theta(t) = v(t),
\end{equation}
where $v(t) \in \mathbb{R}$ is a computational degree of freedom, used to control the progress of the particle along~$Q$.
The function $v(t)$ enables us to later cast the computation of time-optimal motions along reference paths as an OCP, see~\eqref{eq:OCP2}.

The in-homogeneous second-order ordinary differential equation~\eqref{eq:timingLaw} takes care of jumps, but needs to be augmented by conditions on~$\theta$ and~$\dot{\theta}$ at initial time $t=0$ and final time $t=T$. This is done to represent the periodic nature of our content (i.e., the particle reveals the same path many times per second):  
\begin{equation}\label{eq:timingLaw_bc}
	\theta(0)=\theta_0, \quad \theta(T)=\theta_f, \quad \dot{\theta}(0)=\dot{\theta}(T),
\end{equation}
where $\theta_0$ and $\theta_f$ are taken from~\eqref{eq:path} and the total time~$T$ will be optimally determined by the OCP. The first two equations in~\eqref{eq:timingLaw_bc} ensure that the path~$Q$ is fully traversed, while the third one ensures the speeds at the beginning and end of the path match. 

With these additional considerations, we define the (state) vector $\textbf{z}(t):=(\theta(t), \dot{\theta}(t))^\top$, which finally allows us rewrite our second-order differential equations in~\eqref{eq:timingLaw} as a system of first-order differential equations:
\begin{equation} \label{eq:dz_dt}
	\dot{\textbf{z}}(t)=
	\begin{pmatrix}
		0&1\\
		0&0
	\end{pmatrix}\textbf{z}(t)+\begin{pmatrix}
		0\\
		1
	\end{pmatrix}v(t), \quad \textbf{z}(0)=\textbf{z}_0, ~\textbf{z}(T) = \textbf{z}_T.
\end{equation}
Please note that \eqref{eq:dz_dt} is an equivalent \emph{virtual} system to~\eqref{eq:timingLaw} and~\eqref{eq:timingLaw_bc}, solvable using standard Runge-Kutta methods~\cite{butcher2016numerical}. The system~\eqref{eq:dz_dt} is suitable to generate the timing law, but the particle dynamics~\eqref{eq:system_acoustic} still need to be included, as described next.

\subsubsection{Coupling of non-invertible particle dynamics:}\label{sssec:OPT_trap_placement}
To include the particle dynamics~\eqref{eq:system_acoustic} in the computation of the timing, we need to couple them with the system~\eqref{eq:dz_dt}.
The typical approach is to rewrite~\eqref{eq:system_acoustic} as:
\begin{equation}\label{eq:M_implicit}
	M(\ddot{\textbf{p}}(t), {\textbf{p}}(t), {\textbf{u}}(t)):= m\ddot{\textbf{p}}(t) - F(\textbf{p}(t),\textbf{u}(t)) = 0
\end{equation}
and make sure that this equation locally admits an inverse function:
\begin{equation} \label{eq:invModel}
	{\textbf{u}}(t) = M^{-1}(\ddot{\textbf{p}}(t), {\textbf{p}}(t)),
\end{equation}
This would allow us to easily compute the location of our traps.
However, such inversion is not straightforward for our model, and to the best of our knowledge, standard solution methods do not exist for such cases. 

We deal with this challenge by introducing constraints related to our particle dynamics. 
These will allow us to compute feasible timings while delaying the computation of the exact trap location to a later stage in the process.

More specifically, we introduce auxiliary variables~$\zeta_1,...,\zeta_6$ for each trigonometric term in the force~$F$, and we replace $p_i = q_i(\theta)$ for $i \in \{x,y,z\}$ (i.e., particle position exactly matches our reference path):
\begin{subequations}\label{eq:zeta}
	\begin{align}
		\zeta_1 &= \sin\left(\mathcal{V}_{xr}  {\sqrt{(u_x-q_x(\theta))^2+(u_y-q_y(\theta))^2}}\right),\\
		\zeta_2 &= \cos\left(\mathcal{V}_z \cdot {(u_z-q_z(\theta)}\right),\\
		\zeta_3 &= \sin\left(\mathcal{V}_z \cdot {(u_z-q_z(\theta)}\right), \\
		\zeta_4 &= \cos\left(\mathcal{V}_{zr}  {\sqrt{(u_x-q_x(\theta))^2+(u_y-q_y(\theta))^2}}\right),\\
		\zeta_5 &= \sin\phi,\\
		\zeta_6 &= \cos\phi.
	\end{align}
\end{subequations}

With this, we are now able to formally express the force~\eqref{eq:Facou} in terms of $\zeta:=(\zeta_1,...,\zeta_6)$, i.e.,
\begin{equation}
    \tilde{F}(\zeta) := \begin{pmatrix}
    \mathcal{A}_r \zeta_1\zeta_2 \zeta_6 \\
     \mathcal{A}_r \zeta_1\zeta_2 \zeta_5\\
     \mathcal{A}_z \zeta_4\zeta_3
    \end{pmatrix}.
\end{equation}

Using these auxiliary variables, we can now couple the trap-particle dynamics~\eqref{eq:system_acoustic} with the virtual system~\eqref{eq:dz_dt}.
To this end, similar to~\eqref{eq:M_implicit}, we define the following constraint along path~$Q$:
\begin{equation}
	\widetilde M(\theta(t), \dot\theta(t), v(t), \zeta(t)) := m\ddot{\textbf{q}}(\theta(t)) - \tilde{F}({\zeta}(t)) = 0.
\end{equation}

Observe that i) we need~$\dot{\theta}(t)$ due to~\eqref{eq:dpath_dpara_2}; and ii) $\ddot{\theta}(t)$ can be replaced by~$v(t)$ due to~\eqref{eq:timingLaw}. 

Finally, combining this with the prior first-order differential system allows us to conceptually formulate the problem of computing a minimum-time motion along the path~$Q$ as:
\begin{equation}\label{eq:OCP2}
	\begin{aligned}
	    \min_{v,T, \zeta}&~T + \gamma\int_0^T v(t)^2 \mathrm{d}t \\ 
	    \text{subject to}& \\
	    \dot{\textbf{z}}(t)&= \begin{pmatrix}0&1\\0&0\end{pmatrix}\textbf{z}(t)+\begin{pmatrix}
	    0\\1\end{pmatrix}v(t), 
	    \quad \textbf{z}(0) =\textbf{z}_0, ~\textbf{z}(T) = \textbf{z}_T, \\
		0&=\widetilde M(\textbf{z}(t), v(t), \zeta(t)), \\
		\zeta(t)&\in [-1,1]^6.
	\end{aligned}
\end{equation}

The constraint on $\zeta(t)$ is added since the trigonometric structure of~\eqref{eq:zeta} is not directly encoded in the OCP, while $\gamma\geq 0$ is a regularisation parameter.

\begin{figure}[tb]
  \includegraphics[width=1\columnwidth]{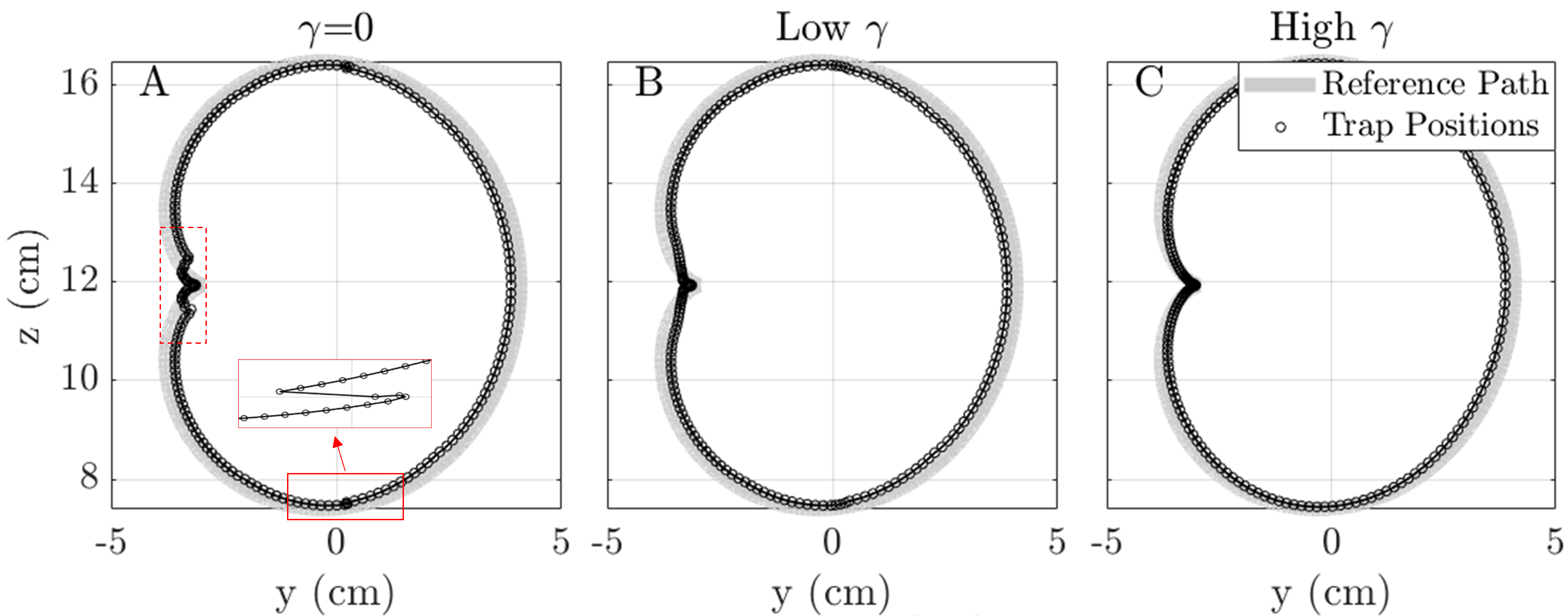}
 \caption{Effect of the regularisation parameter $\gamma$ on the position of the optimised traps.
 }
\label{fig:gamma_comparison}
\Description{Three subplots A, B and C show a cardioid shape in two dimensional space. The x-axis shows the horizontal spatial component from -5 to 5 centimetres, at steps of 5, while the y-axis shows the vertical from 8 to 16 centimetres, at steps of 4. Subplot A shows the computed trap positions when the regularisation parameter gamma is low. Sudden jumps at the highest and lowest vertical position, as well as at the sharp edge of the shape are highlighted.}
\end{figure}

The case~$\gamma=0$ corresponds to strictly minimising time, which typically leads to an aggressive use of the forces around the trap. The example in Figure \ref{fig:gamma_comparison}(A) shows the trap accelerating the particle before arriving at the corner, then applying aggressive deceleration before it reaches the corner. 
At the top and bottom of the shape, again the particle is accelerated strongly, then it is suddenly decelerated by the traps placed behind it, in order to move around the curve.
This would be the optimum solution in an ideal case (i.e., a device working \emph{exactly} as per our model), but inaccuracies in the real device can make them unstable. Moreover, we observed the overall reductions in rendering time to usually be quite small. 

The regularisation ($\gamma>0$) enables \emph{nearly} time-optimal solutions. More specifically, this is a strictly convex regularisation that penalises high magnitudes of the virtual input $v = \ddot{\theta}$. This is most closely related to the accelerations applied to the particle, hence avoiding aggressive acceleration/deceleration as shown in Figures~\ref{fig:gamma_comparison}(B) and (C). Several heuristics could be proposed to automate selection of a suitable value for $\gamma$ (e.g., use smallest $\gamma$ ensuring that the dot product of $\dot{\textbf{p}}(t)$ and $\dot{\textbf{u}}(t)$ remains always positive), but this step is considered beyond the scope of the current paper.   

Note that the OCP in~\eqref{eq:OCP2} yields the (nearly) optimal timing along the path~$Q$, respecting the particle dynamics and hence solving \emph{Challenge 1}.
More\-over, it yields the required forces through $\zeta(t)$. 
However, it does not give the trap trajectory~$\textbf{u}(t)$ that generates these forces.
To obtain the trap trajectory and thus solve \emph{Challenge~2}, we refine~\eqref{eq:OCP2} in the following section.

\subsection{Computing the Trap Trajectory}\label{ssec:extract-trap-trajectory}
The second stage in our approach deals with \emph{Challenge 2}, computing the trap trajectory~$\textbf{u}(t)$ based on the solution of~\eqref{eq:OCP2}. In theory, we would use the values for $\zeta_i(t)$, $i=1,...,6$ and $q_j(\theta(t))$, $j\in\{x,y,z\}$ to determine the particle position and force required at each moment in time, obtaining the required trap position~$\textbf{u}(t)$ by solving~\eqref{eq:zeta}. 

In practice, solving~$\textbf{u}(t)$ from~\eqref{eq:zeta} numerically is not trivial, particularly for values of $\zeta_i$ close to $\pm 1$, where numerical instabilities could occur, particularly for $\zeta_1$ and $\zeta_4$. 

We attenuate these difficulties by providing further structure to the OCP~\eqref{eq:OCP2}, specifically regarding~$\zeta$: 
\begin{equation}\label{eq:zeta_trig_pythagoras}
	\zeta_2^2+\zeta_3^2 = 1, \quad \zeta_5^2+\zeta_6^2 = 1.
\end{equation}

We also constrain the solvability of~\eqref{eq:zeta} by using a constant back-off $\varepsilon\in\interval[open]{0}{1}$ in order to avoid numerical instabilities:
\begin{equation}\label{eq:consistConstraints}
    -1+\varepsilon \leq \zeta_i \leq 1-\varepsilon, \quad i=1,...,6.
\end{equation}
 
For the sake of compact notation, we summarise the above constraints in the following set notation
\begin{equation}
    \mathcal{Z} :=  \left\{\zeta \in  \mathbb{R}^6 ~ | ~\eqref{eq:zeta_trig_pythagoras} \text{ and } \eqref{eq:consistConstraints}
    \text{ are satisfied}\right\}.
\end{equation}

The final OCP is then given by
\begin{equation}\label{eq:OCP3}
	\begin{aligned}
	    \min_{v,T, \zeta}&~T + \gamma\int_0^T v(t)^2 \mathrm{d}t \\ 
	    \text{subject to}& \\
	    \dot{\textbf{z}}(t)&= \begin{pmatrix}0&1\\0&0\end{pmatrix}\textbf{z}(t)+\begin{pmatrix}
	    0\\1\end{pmatrix}v(t), 
	    \quad \textbf{z}(0) =\textbf{z}_0, ~\textbf{z}(T) = \textbf{z}_T, \\
	   0&=\widetilde M(\textbf{z}(t), v(t), \zeta(t)), \\
	   \zeta(t)&\in \mathcal{Z}.
	\end{aligned}
\end{equation}

Finally, given $\zeta(t) \in \mathcal{Z}$ and $\theta(t)$ from the solution of~\eqref{eq:OCP3}, we numerically solve~\eqref{eq:zeta} for $u_i$, $i\in\{x,y,z\}$, using Powell's dog leg method~\cite{Mangasarian94}. Please note that higher values of $\varepsilon$ will limit the magnitudes of the forces exploited by our approach. A simple solution is to perform a linear search over $\varepsilon$, only increasing its value and recomputing the solution if Powell's method fails to converge to a feasible trap location.  
For details on discretising the OCP~\eqref{eq:OCP3}, we refer to the Supplementary Material S3.

%-----------------------------------------------------------------------------

\section{Evaluation}
\label{sec:eval}
This section evaluates the capability of our algorithm to support content creation by generating physically feasible trap trajectories given only a reference path (i.e., a shape) as an input. We select a range of shapes and analyse the effects each step in our approach has on the final achievable size, rendering frequency, and reconstruction error, for each shape. 
We also inspect how these steps influence the presence of visual distortions in the end result.
Finally, we analyse the resulting acceleration profiles.

\begin{figure}[b]
  \includegraphics[width=\columnwidth]{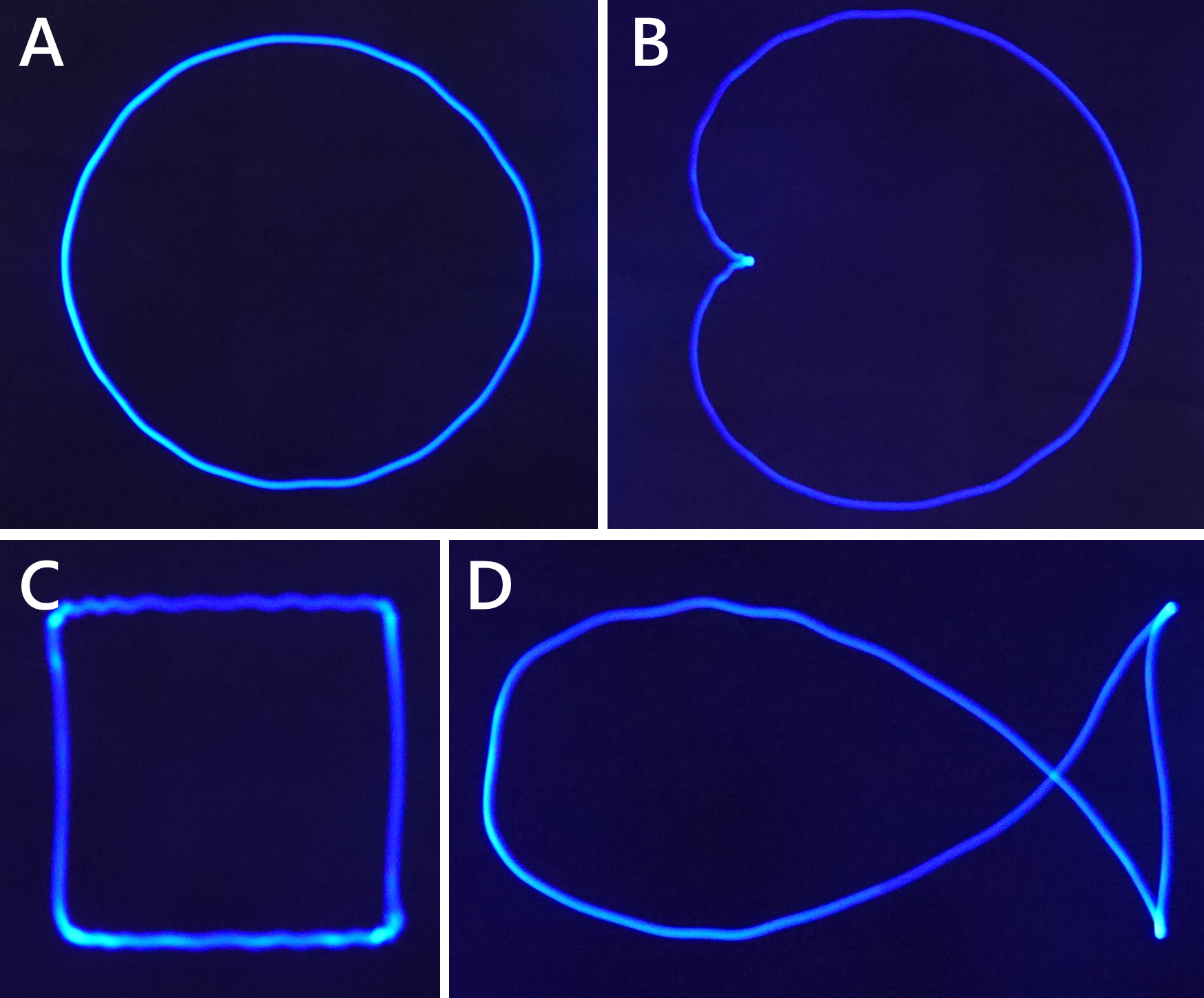}
  \caption{Test shapes used for evaluation, as rendered by our \emph{OptiTrap} approach. The circle (A) provides a trivial timing case. The cardioid (B) and squircle (C) represent simple shapes featuring sharp corners and straight lines. The fish (D) is used as an example of composite shape featuring corners, straight lines, and curves.}
  \label{fig:EvalShapes}
\Description{The image consists of four photos A, B, C and D. The photos show illuminated outlines of the different shapes rendered by the levitation interface in mid-air.}
\end{figure}

\subsection{Test Shapes}
\label{ssec:eval_test_shapes}

We used four test shapes for the evaluation: a circle, a cardioid, a squircle, and a fish, all shown in Figure~\ref{fig:EvalShapes}.

The circle is selected as a trivial case in terms of timing. It can be optimally sampled by using a constant angular speed (i.e., constant acceleration), and is there to test whether our solution converges towards such optimum solutions. The cardioid and squircle represent simple shapes featuring sharp corners (cardioid) and straight lines (squircle), both of them challenging features. Finally, the fish is selected as an example composite shape, featuring all elements (i.e., corners, straight lines, and curves). 

All of these shapes can be parameterised by low-frequency sinusoids (see Supplementary Material S4), ensuring the path parameter~$\theta$ progresses slowly in areas of high curvature. 
This provides a good starting point for our baseline comparison that we explain in Section~\ref{ssec:eval_conditions}. 
While our approach works for content in 3D, we perform the evaluation on shapes in a 2D plane of the 3D space to facilitate the analysis of accelerations in Section~\ref{ssec:eval_accelerations}, which we will perform in terms of horizontal and vertical accelerations (i.e., the independent factors given our axis-symmetric model of forces), allowing us to validate \emph{OptiTrap}'s awareness of particle directions and how close it can get to the maximum accelerations allowed by the dynamics of acoustic traps, as discussed in Section~\ref{ssec:model-acoustic-forces}.

Notice in Figure~\ref{fig:EvalShapes}, that due to the timing differences, different shape elements exhibit different levels of brightness. 
To obtain homogeneous brightness along the shape, please refer to the solution proposed by \cite{Hirayama19}, where the particle illumination is adjusted to the particle speed.

\subsection{Conditions Compared}
\label{ssec:eval_conditions}

We compare three approaches to rendering levitated shapes, where each approach subsequently addresses one of the challenges introduced in Section~\ref{sec:prob_stat}. 
This will help us assess the impact that each challenge has on the final results obtained.

The first condition is a straight-forward \emph{Baseline}, with homogeneous sampling of the path parameter, which matches the example strategy shown in Figure~\ref{fig:timings}(B), and placing traps where the particle should be. The \emph{Baseline} still does not address any of the challenges identified (i.e., optimum timing or trap placement), but it matches approaches used in previous works \cite{Hirayama19, Plasencia20, FushimiLimits} and will illustrate their dependence on the specific shape and initial parametrisation used.

The second condition, \emph{OCP\_Timing}, makes use of our \emph{OptiTrap} approach to compute optimum and feasible timing (see Subsection~\ref{sssec:OPT_timing}), but still ignores \emph{Challenge 2}, assuming that particles match trap location (i.e., it skips Subsection~\ref{sssec:OPT_trap_placement}). The third condition is the full \emph{OptiTrap} approach, which considers feasibility and trap-particle dynamics and deals with both the timing and trap placement challenges. 

These conditions are used in a range of comparisons involving size, frequency, reconstruction error as well as comparative analysis of the effects of each condition on the resulting particle motion, detailed in the following sections. 

\subsection{Maximum Achievable Sizes}
\label{ssec:eval_sizes}
This section focuses on the maximum sizes that can be achieved for each test shape and condition, while retaining an overall rendering time of 100ms\footnote{The circle rendered at 100ms would exceed the size of our device's working volume. Thus, this shape was rendered at 67ms.} (i.e., PoV threshold). For each of these cases, we report the maximum achievable size and reliability, which were determined as follows.

Maximum size is reported in terms of shape width and in terms of meters of \emph{content per second} rendered \cite{Plasencia20}, and their determination was connected to the feasibility of the trap trajectories, particularly for the \emph{Baseline} condition. 

\begin{table*}
\caption{Maximum shape width and meters of content per second achieved with the \emph{Baseline}, \emph{OCP\_Timing}, and \emph{OptiTrap} approach at constant rendering frequency. Successful trials out of 10.}
\centering
\begin{tabular}{ c|c|c|c|c|c|c|c|c|c|c|c|} 
 \cline{3-12}
\multicolumn{2}{c}{}&\multicolumn{3}{|c}{\emph{Baseline}}& \multicolumn{3}{|c}{\emph{OCP\_Timing}}&\multicolumn{3}{|c|}{\emph{OptiTrap}}&\multicolumn{1}{c|}{Increase  }\\
 \cline{1-11}
\multicolumn{1}{|l|}{Shape}  &Freq.& Width  & Content per & Success & Width  & Content per& Success & Width & Content per& Success & in Width w.r.t.\\ 
 \multicolumn{1}{|l|}{}  & (Hz) &  (cm) & Second (m)&  Rate &  (cm)  &  Second (m)&  Rate& (cm) & Second (m)& Rate &the \emph{Baseline}\\ 
 \hline 
\multicolumn{1}{|l|}{Circle} & 15& 7.00& 3.30 & 10/10&7.00 & 3.30 & 10/10  &7.00 & 3.30 &  10/10 & 0\% \\ 
 \hline 
\multicolumn{1}{|l|}{Cardioid} &10 & 8.05 & 2.48& 10/10& 9.09& 2.80& 10/10  & 9.09 &2.80 & 10/10 &12.9\% \\ 
 \hline 
\multicolumn{1}{|l|}{Squircle} &10&  0.80&   0.29 & 9/10&5.30& 1.90 & 10/10 & 5.30& 1.90 &  10/10 & 562.5\%\\ 
 \hline 
 \multicolumn{1}{|l|}{Fish} &10&  7.77& 2.42 & 10/10&8.76 & 2.75 &10/10  &8.76& 2.75 & 10/10 & 12.7\% \\ 
 \hline 
\end{tabular}
\label{table:shape_sizes_results}
\end{table*}

More specifically, we determined maximum feasible sizes for the \emph{Baseline} condition by conducting an iterative search. That is, we increased the size at each step and tested the reliability of the resulting shape. We considered the shape feasible if it could be successfully rendered at least 9 out of 10 times in the actual levitator. On a success, we would increase size by increasing the shape width by half a centimetre. On a failure, we would perform a binary search between the smallest size failure and the largest size success, stopping after two consecutive failures and reporting the largest successful size. This illustrates the kind of trial and error that a content designer would need to go through without our approach and it was the most time consuming part of this evaluation.

For the other two conditions (i.e., \emph{OCP\_Timing} and \emph{OptiTrap}), maximum sizes could be determined without needing to validate their feasibility in the final device. Again, we iteratively increased the shape size using the same search criteria as before (i.e., 5mm increases, binary search). We assumed any result provided would be feasible (which is a reasonable assumption; see below) and checked the resulting total rendering time, increasing the size if the time was still less than 100ms. At each step, a linear search was used, iteratively increasing $\varepsilon$, until feasible trap locations could be found (i.e., equation~\eqref{eq:zeta} could be solved without numerical instabilities). Adjusting the value of the regularisation parameter $\gamma$ was done by visually inspecting the resulting trap trajectories, until no discontinuities could be observed (see Figure \ref{fig:gamma_comparison}). Please note that this is a simple and quick task, which, unlike the \emph{Baseline} condition, does not involve actual testing on the levitation device and could even be automated. All solutions provided can be assumed feasible, and the designer only needs to choose the one that better fits their needs.  

Once the maximum achievable sizes were determined, these were tested in the actual device. We determined the feasibility of each shape and condition by conducting ten tests and reporting the number of cases where the particle succeeded to reveal the shape. This included the particle accelerating from rest, traversing/revealing the shape for 6 seconds and returning to rest.

Table \ref{table:shape_sizes_results} summarises the results achieved for each test shape and condition. First of all, it is worth noting that all trials were successful for  \emph{OptiTrap} and \emph{OCP\_Timing} approaches, confirming our assumption that the resulting paths are indeed feasible and underpinning \emph{OptiTrap}'s ability to avoid trial and error on the actual device during content creation. 

The results and relative improvements in terms of size vary according to the particular condition and shape considered. For instance, it is interesting to see that all conditions yield similar final sizes for the circle, showing that \emph{OptiTrap} (and \emph{OCP\_Timing}) indeed converge towards optimum solutions.

More complex shapes where the \emph{Baseline} parametrisation is not optimal (i.e., cardioid, squircle, and fish), show increases in size when using \emph{OCP\_Timing} and \emph{OptiTrap}. More interestingly, the increases in size vary greatly between shapes, showing increases of around $12\%$ for the fish and cardioid, and up to $562\%$ for the squircle. This is the result of the explicit parametrisation used, with reduced speeds at corners in the fish and cardioid cases, but not in the case of the squircle. 

It is also interesting to see that  \emph{OCP\_Timing} and \emph{OptiTrap} maintain high values of \emph{content per second} rendered, independently of the shape. That is, while the performance of the \emph{Baseline} approach is heavily determined by the specific shape (and parametrisation) used, the OCP-based approaches (\emph{OCP\_Timing} and \emph{OptiTrap}) yield results with consistent \emph{content per second}, determined by the capabilities of the device (but not so much by the specific shape). Finally, please note that no changes in terms of maximum size can be observed between \emph{OCP\_Timing} and \emph{OptiTrap}.

\subsection{Maximum Rendering Frequencies}
\label{ssec:eval_frequencies}

\begin{table}[b]
\caption{Maximum rendering frequencies achieved with the \emph{Baseline}, \emph{OCP\_Timing}, and \emph{OptiTrap} approach, for shapes of equal size. Successful trials out of 10.}
\centering
\begin{tabular}{ c|c|c|c|c|c|c|} 
 \cline{2-7}
&\multicolumn{2}{c|}{\emph{Baseline}}&\multicolumn{2}{c}{\emph{OCP\_Timing}}&\multicolumn{2}{|c|}{\emph{OptiTrap}}\\ 
 \cline{1-7}
\multicolumn{1}{|l|}{Shape}& Freq. &  Success & Freq. & Success &Freq. &  Success \\ 
\multicolumn{1}{|l|}{} & (Hz)& Rate& (Hz)& Rate& (Hz)& Rate  \\ 
 \hline 
\multicolumn{1}{|l|}{Circle} &15 & 10/10 & 15 &10/10 & 15  & 10/10 \\ 
 \hline 
\multicolumn{1}{|l|}{Cardioid}& 8  & 10/10 & 10 &10/10 &10 & 10/10 \\ 
 \hline 
\multicolumn{1}{|l|}{Squircle} & 4 & 10/10 & 10 &10/10 &10 & 10/10 \\ 
 \hline 
 \multicolumn{1}{|l|}{Fish} & 9 & 9/10& 10 &10/10  &10 & 10/10\\ 
 \hline 
\end{tabular}
\label{table:shape_frequency_results}
\end{table}

In this second evaluation, we assessed the effect of the timing strategy on the maximum achievable rendering frequencies. To do this, we selected the maximum achievable sizes obtained in the prior evaluation for each shape. We then reproduced such sizes with the \emph{Baseline} approach, searching for the maximum frequency at which this approach could reliably render the shape (using the same searching and acceptance criteria as above). 

That is, while we knew \emph{OCP\_Timing} and \emph{OptiTrap} could provide 10Hz for these shapes and sizes, we wanted to determine the maximum frequency at which the \emph{Baseline} would render them, as to characterise the benefits provided by optimising the timing.

The results are summarised in Table \ref{table:shape_frequency_results}. As expected, no changes are produced for the circle. However, there is a 25\% increase for the cardioid, 150\% for the squircle, and 11\% increase for the fish using the \emph{OptiTrap} method. Again, please note that no differences can be observed in terms of maximum frequency between \emph{OCP\_Timing} and \emph{OptiTrap}. 

\subsection{Reconstruction Accuracy}
\label{ssec:eval_accuracy}

To evaluate the reconstruction accuracy, we recorded each of the trials in our maximum size evaluations (see Section~\ref{ssec:eval_sizes}) using an OptiTrack camera system. We then computed the Root Mean Squared Error (RMSE) between the recorded data and the intended target shape, taking 2s of shape rendering into account (exclusively during the cyclic part, ignoring ramp-up/ramp-down). For a fair comparison across trials, we normalise the RMSE with respect to the traversed paths, by dividing the RMSE by the total path length of each individual test shape. The results are summarised in Table~\ref{table:RMSE_results}.

\begin{table}[b]
\caption{RMSE and Path-normalised (PN) RMSE with respect to the total path length of each test shape, for the \emph{Baseline}, \emph{OCP\_Timing}, and \emph{OptiTrap} approach, at constant rendering frequency.}
\centering
\begin{tabular}{ c|c|c|c|c|c|c|} 
 \cline{2-7}
&\multicolumn{2}{|c}{\emph{Baseline}}&
\multicolumn{2}{|c}{\emph{OCP\_Timing}}&
\multicolumn{2}{|c|}{\emph{OptiTrap}}\\
 \hline 
\multicolumn{1}{|l|}{Shape}  &  RMSE & PN &  RMSE & PN & RMSE & PN\\ 
\multicolumn{1}{|l|}{}  &  (cm) & RMSE & (cm) & RMSE & (cm) & RMSE\\ 
 \hline 
\multicolumn{1}{|l|}{Circle} &  0.196 & 0.893 & 0.146 & 0.666& 0.114 &  0.521\\ 
 \hline 
\multicolumn{1}{|l|}{Cardioid} & 0.142 & 0.575 & 0.153 & 0.545 & 0.120 & 0.429\\ 
 \hline 
\multicolumn{1}{|l|}{Squircle} & 0.056 & 1.956& 0.0824 &0.434 &  0.080 & 0.421\\ 
 \hline 
 \multicolumn{1}{|l|}{Fish} & 0.111 &0.458 & 0.147&0.536 &0.0581 & 0.212\\ 
 \hline 
\end{tabular}
\label{table:RMSE_results}
\end{table}

On average, both \emph{OCP\_Timing} and \emph{OptiTrap} provide better results in terms of accuracy, when compared to the \emph{Baseline}. 
It is particularly worth noting that the accuracy results, in terms of the raw RMSE for \emph{OptiTrap} and \emph{OCP\_Timing}, are better for the cardioid when compared to the \emph{Baseline}, even though a larger shape is being rendered. 
While the raw RMSE of \emph{OCP\_Timing} and \emph{OptiTrap} is slightly larger for the squircle, we need to take into account that the size rendered by \emph{OCP\_Timing} and \emph{OptiTrap} is almost 6 times larger. 
Comparing accuracy in terms of normalised RMSE shows consistent increases of accuracy for \emph{OptiTrap} compared to the \emph{Baseline}, with overall decreases in the RMSE of $41.7\%$ (circle), $25.4\%$ (cardioid), $78.5\%$ (squircle), and $53.8\%$ (fish). 
\emph{OCP\_Timing} shows a decrease of $25.4\%$, $5.14\%$, and $77.8\%$ of the normalised RMSE for the circle, cardioid, and squircle, with the exception of the fish, where the normalised RMSE increased by $17.1\%$, when compared to the \emph{Baseline}.

\begin{figure}[tb]
  \includegraphics[width=\columnwidth]{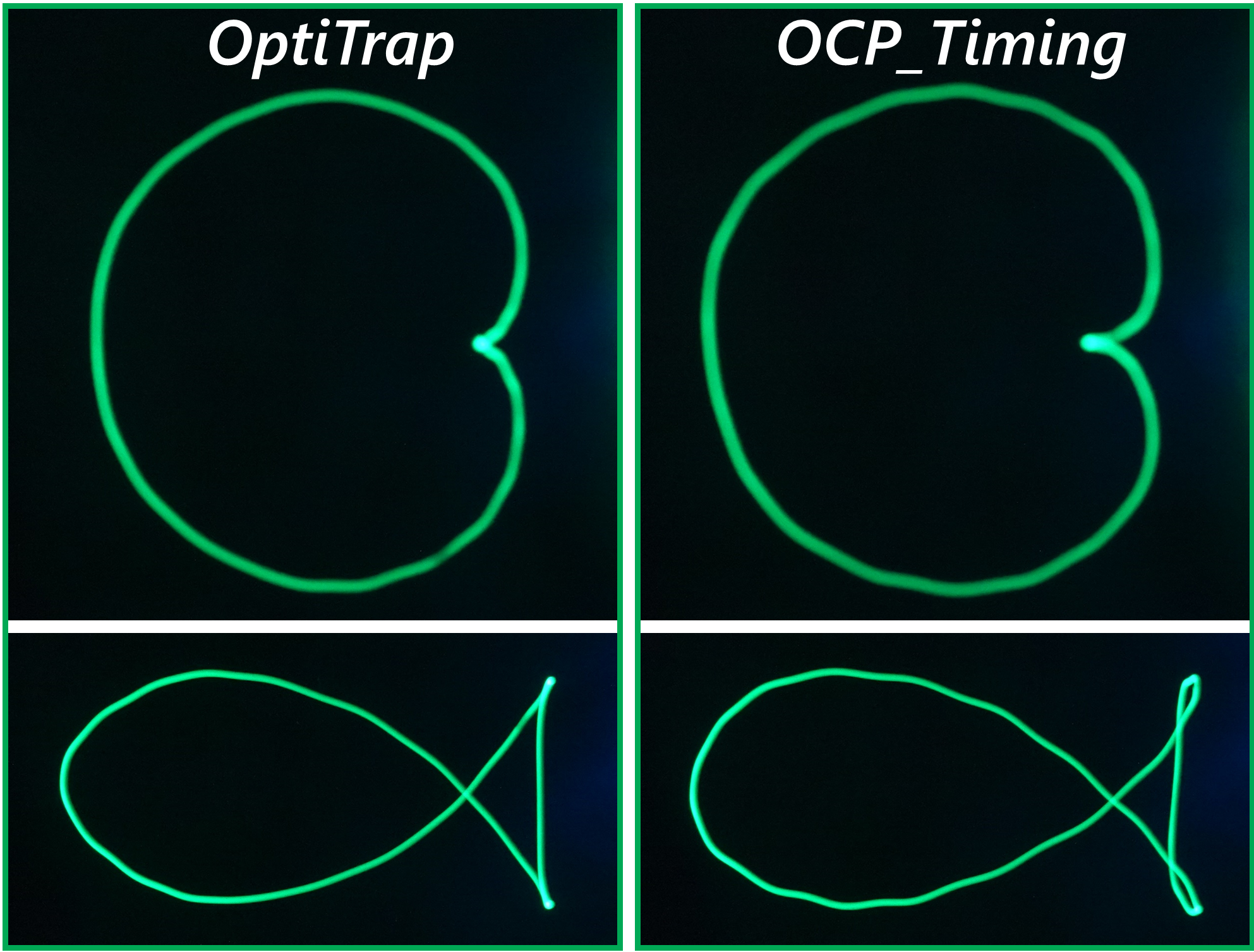}
  \caption{Visual comparison between shapes rendered with traps located according to particle dynamics using \emph{OptiTrap} (left) and traps placed along the reference path using \emph{OCP\_Timing} (right) for the cardioid (top) and the fish (bottom) test shapes. Please note undesired increases in size and error in sharp features, such as corners.}
  \label{fig:eval_trap_location}
\Description{The image consists of a two-by-two photo matrix. There are two photos of a rendered cardioid in the first row, and fish in the second row. The cardioid and the fish in the first column are rendered using the OptiTrap algorithm, and in the second column the two shapes are rendered using OCP_Timing. The levitated graphics in the second column are a bit larger in size and less precise than those in the first column. }
\end{figure}

Comparing the reconstruction accuracy of \emph{OCP\_Timing} and \emph{OptiTrap} highlights the relevance of considering trap dynamics to determine trap locations (i.e., Section~\ref{ssec:extract-trap-trajectory}). \emph{OptiTrap} consistently provides smaller RMSE than \emph{OCP\_Timing},
as a result of considering and accounting for the trap-to-particle displacements required to apply specific accelerations. 
We obtain a $21.8\%$, $21.3\%$, $2.9\%$ and $60.5\%$ decrease in the normalised RMSE, for the circle, cardioid, squircle, and fish, respectively. 
Such differences are visually illustrated in Figure \ref{fig:eval_trap_location}, showing the effects on the cardioid and fish, for the \emph{OCP\_Timing} and \emph{OptiTrap} approaches. It is worth noting how placing the traps along the reference path results in the cardioid being horizontally stretched, as the particle needs to retain larger distances to the trap to keep the required acceleration (horizontal forces are weaker than vertical ones). This also results in overshooting of the particle at corner locations, which can be easily observed at the corner of the cardioid and in the fins of the fish. 
For completeness, the visual comparison for the remaining two test shapes is provided in the Supplementary Material S5.

\subsection{Analysis of Acceleration profiles}
\label{ssec:eval_accelerations}

\begin{figure}[b]
  \includegraphics[width=1\columnwidth]{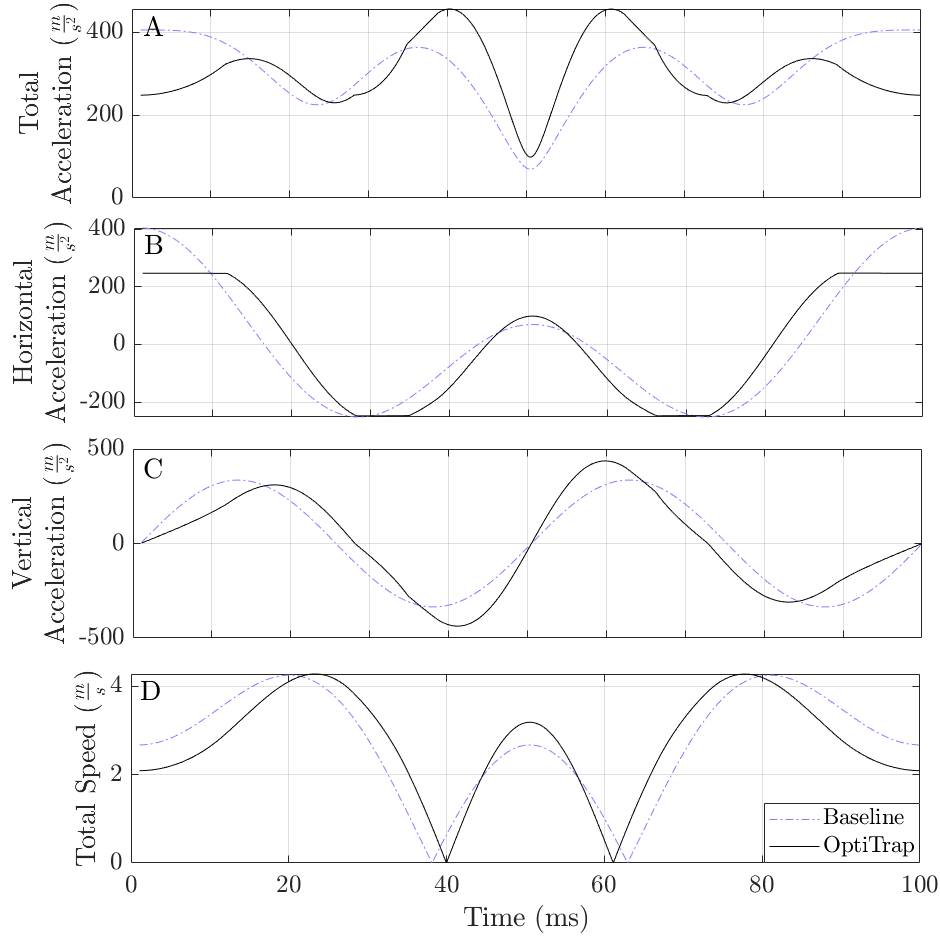}
  \caption{Particle acceleration and speed for the fish rendered at 10~Hz using \emph{OptiTrap} (solid black) and the \emph{Baseline} (dash-dotted purple). While the speed profiles (D) of both trap trajectories are similar, the trap trajectory generated by our approach is feasible, whereas the one generated by the \emph{Baseline} is not. The main reason is that the \emph{Baseline} exceeds the feasible horizontal acceleration, while our approach caps it to feasible values (B). Our approach compensates by using higher (available) vertical acceleration (C). Note that the times where the \emph{Baseline} applies a higher total acceleration (A) than our approach are those where our approach respects the constraints on the feasible horizontal acceleration (B).}
  \label{fig:force_plots}
\Description{The figure consists of four subplots. Subplot A shows the total acceleration in metres per seconds squared from 0 to 400, in steps of 200.
Subplot B shows the horizontal acceleration in metres per seconds squared from -200 to 400, in steps of 200.
Subplot C shows the vertical acceleration in metres per seconds squared from -500 to 500, in steps of 500.
Subplot D shows the total speed in metres per seconds from 0 to 4, in steps of 2. All quantities are plotted as functions of time, going from 0 to 100 milliseconds, at steps of 20.}
\end{figure}

Finally, we examined the acceleration profiles produced by \emph{OptiTrap} and how these differ from the pre-determined acceleration profiles of the \emph{Baseline}. Please note that we only compare and discuss \emph{OptiTrap} and \emph{Baseline} in Figure \ref{fig:force_plots}, and only for the fish shape. \emph{OCP\_Timing} is not included, as it results in similar acceleration profiles as \emph{OptiTrap}. Only the fish is discussed, as it already allows us to describe the key observations that can be derived from our analysis.
For completeness, the acceleration profiles for all remaining test shapes are included in the Supplementary Material S6. 

As introduced above, Figure \ref{fig:force_plots} shows the particle accelerations and speeds of a particle revealing a fish shape of maximum size (i.e., a width of 8.76cm), rendered over 100ms using the \emph{OptiTrap} and \emph{Baseline} approaches. It is worth noting that while \emph{OptiTrap} succeeded in rendering this shape, \emph{Baseline} did not (the maximum width for \emph{Baseline} was 7.77cm).

A first interesting observation is that although \emph{OptiTrap} provides lower total accelerations than the \emph{Baseline} during some parts of the path (see Figure \ref{fig:force_plots}(A)), it still manages to reveal the shape in the same time. 
The key observation here is that the regions where the total acceleration is lower for \emph{OptiTrap} match with the parts of the path where the horizontal acceleration is very close to its maximum; for example, note the flat regions in Figure~\ref{fig:force_plots}(B) of around $\pm 300 m/{s}^2$). This is an example of \emph{OptiTrap}'s awareness of the dynamics and capabilities of the actual device, limiting the acceleration applied as to retain the feasibility of the path. 

Second, it is worth noting that maximum horizontal accelerations are significantly smaller than vertical accelerations. As such, it does make sense for horizontal displacements to become the limiting factor. In any case, neither acceleration exceeds its respective maximum value. The \emph{OptiTrap} approach recovers any missing time by better exploiting areas where acceleration is unnecessarily small. It is also worth noting that the value of the horizontal and vertical accelerations are not simply being capped to a maximum acceleration value per direction (e.g., simultaneously maxing out at $\pm 300 m/{s}^2$ and $\pm 600 m/{s}^2$ in the horizontal and vertical directions), as such cases are not feasible according to the dynamics of our acoustic traps.

Third, it is interesting to note that the final acceleration profile is complex, retaining little resemblance with the initial parametrisation used. This is not exclusive for this shape and can be observed even in relatively simple shapes, such as the cardioid in Figure~\ref{fig:timings}(D) or the remaining shapes, provided in the Supplementary Material S6. 

The complexity of these profiles is even more striking if we look at the final speeds resulting from both approaches (see Figure \ref{fig:force_plots}(D)). Even if the acceleration profiles of \emph{Baseline} and \emph{OptiTrap} are very different, their velocity profiles show only relatively subtle differences. But even if these differences are subtle, they mark the difference between a feasible path (i.e., \emph{OptiTrap}) and a failed one (i.e., \emph{Baseline}). These complex yet subtle differences also illustrate how it is simply not sensible to expect content designers to deal with such complexity, and how \emph{OptiTrap} is a necessary tool to enable effective exploitation of PoV levitated content.  

%-----------------------------------------------------------------------------
\section{Discussion}
This paper presented \emph{OptiTrap}, an automated approach for optimising timings and trap placements, as to achieve feasible target shapes. We believe this is a particularly relevant step for the adoption of levitation PoV displays, as it allows the content creator to focus on the shapes to present, with feasible solutions being computed automatically, while making effective usage of the capabilities of the device. As such, we hope \emph{OptiTrap} to become an instrumental tool in helping explore the actual potential of these displays.

However, \emph{OptiTrap} is far from a complete content creation tool. Such a tool should consider the artist's workflows and practices. Similarly, testing and identifying most useful heuristics to tune our approach (e.g., the regularisation) or visualisation tools identifying tricky parts of the shapes (i.e., requiring high accelerations) should be included as a part of this process. Our goal is simply to provide the base approach enabling this kind of tools.

Even this base approach can be extended in a variety of ways. 
Our levitator prototype is built from off-the-shelf hardware, and is still subject to inaccuracies that result in distortions in the sound-fields generated \cite{Fushimi19}. As such, more accurate levitation hardware or, alternatively, a model reflecting the dynamics of the system in a more accurate manner would be the most obvious pathway to improve our approach. It is worth noting that both factors should be advanced jointly. A more accurate model could also be less numerically stable, potentially leading to worse results if the hardware is not accurate enough. 

The high update rates of 10kHz required by the levitator and the millisecond delays introduced by optical tracking systems indicate that closed-loop approaches can be both promising and challenging avenues to explore. 
Assuming a tracking device synchronised with the levitation device (as to map current trap locations with real particle positions), \emph{OptiTrap} could be combined with learning-based approaches. These learning-based approaches will require example paths (i.e., with initial timing and trap placement), and their achievable complexity and convergence will be limited by the examples provided. 
In such cases, \emph{OptiTrap} can be used to always generate feasible initial trajectories (i.e., to avoid system restarts on failure).
Thus learning-based approaches could be used to further refine \emph{OptiTrap} beyond our current model, as to account for device inaccuracies such as those discussed by~\cite{Fushimi19}.

However, higher gains can be obtained from more radical changes in the approach. For instance, our approach optimises the timing and trap placement, but it does not modify the target shapes. As illustrated in Figure~\ref{fig:force_plots}, slight changes in speed have significant effects on the acceleration profiles (and feasibility) of the shapes. This is even more prominent for position, where small changes can heavily influence the acceleration and feasibility of target shapes. As such, approaches exploiting subtle modifications to the shape could lead to significant gains in rendering performance.

Another interesting possibility would be extending our approach to use several particles. As shown in \cite{Plasencia20}, while the use of several particles does not increase the overall power that can be leveraged, it does allow for increased flexibility. That is, the intensity/forces of each trap can be individually and dynamically adjusted, as to match the needs of the region of the path that each particle is revealing. Also, particles can each be rendering specific independent features, so the content is not limited to a single connected path, and the particles do not waste time/accelerations traversing parts of the path that will not be illuminated (i.e., visible). 
This approach, however, entails significant challenges. The first obvious challenge is the reliability of the intensity control of the traps. \cite{Plasencia20} demonstrate accurate control of the stiffness at the centre of the traps, but the effects of multiple (interfering) traps in each trap's topology (i.e., how forces distribute around the trap) is yet to be studied. A second challenge is that each particle is not forced to traverse the path at the same speed/rates, with such independent timing progression becoming an additional degree of freedom to account for. 

Finally, further extensions to our work can come from its application to domains other than PoV displays. An obvious next step would be to adapt \emph{OptiTrap} to photophoretic displays, which trap particles using optical traps instead of acoustic traps \cite{Smalley18,Kumagai21}. This would involve including a model of the dynamics of such optical traps, but it can also involve further challenges, such as modelling the response times of galvanometers and LC panels involved in creating the trap. 

Our approach can also be adapted to applications requiring objects to be transported quickly and accurately. For example, contactless transportation of matter has a wealth of applications in areas such as the study of physical phenomena, biochemical processes, materials processing, or pharmaceutics \cite{Foresti12549}. Our method can help solving such problems by directly computing a rest-to-rest solution of the matter to be transported, given an estimation of the mass of that matter, the acoustic force acting on the matter, and a path from start to target position. 

%-----------------------------------------------------------------------------
\section{Conclusion}
In this paper we proposed the first structured numerical approach to compute trap trajectories for acoustic levitation displays. 
\emph{OptiTrap} automatically computes physically feasible and nearly time-optimal trap trajectories to reveal generic mid-air shapes, given only a reference path. 
Building on a novel multi-dimensional approximation of the acoustic forces around the trap, we formulate and show how to solve a non-linear path following problem without requiring or exploiting differential flatness  of the system dynamics.
We demonstrate increases of up to 563\% in size and up to 150\% in frequency for several shapes. Additionally, we obtain better reconstruction accuracy with up to a 79\% decrease in the path-normalised RMSE.
While previously, feasible trap trajectories needed to be tuned manually for each shape and levitator, our approach requires calibration of each individual levitator just once. 
We are confident that the ideas in this paper could form the basis for future content authoring tools for acoustic levitation displays and bring them a key step closer to real-world applications. 
%-----------------------------------------------------------------------------
\begin{acks}
This research has received funding from the European Union’s Horizon 2020 research and innovation programme under grant agreement \#737087 (Levitate) and from the AHRC UK-China Research-Industry Creative Partnerships (AH/T01136X/2).
\end{acks}

\bibliographystyle{ACM-Reference-Format}
\bibliography{OptiTrap-base}

\end{document}